\documentclass[%
 reprint,
preprintnumbers,
 amsmath,amssymb,
 aps,
]{revtex4-1}

\usepackage[dvipsnames]{xcolor}
\usepackage{graphicx}% Include figure files
\usepackage{dcolumn}% Align table columns on decimal point
\usepackage{bm}% bold math
\usepackage{hyperref}% add hypertext capabilities
\usepackage{multirow}

\definecolor{myblue}{rgb}{0.152941176,0.549019608,0.670588235}
\definecolor{newred}{cmyk}{0,1,1,0.2}
\definecolor{newblue}{cmyk}{1,1,0,0.1}
\hypersetup{
    pdftoolbar=true,        
    pdfmenubar=true,        
    pdffitwindow=false,     
    pdfstartview={FitH},    
    pdftitle={On the Nonzero Lepton Asymmetry in the Early Universe},
    pdfauthor={Anne-Katherine Burns, Tim Tait and Mauro Valli},
    pdfkeywords={BBN} {lepton asymmetry} {Neff} {Early Universe},
    pdfnewwindow=true,
    colorlinks=true,
    linkcolor=myblue,
    citecolor=myblue,
    filecolor=myblue,
    urlcolor=myblue,
    linktocpage=true
}

\def\equationautorefname~#1\null{Eq.\,(#1)\null}
\newcommand{\appendixref}[1]{\hyperref[#1]{appendix~\ref{#1}}}

\begin{document}

\preprint{UCI-TR-2022-10}
\preprint{YITP-SB-2022-22}

\title{Indications for a Nonzero Lepton Asymmetry from Extremely Metal-Poor Galaxies}

\author{Anne-Katherine Burns}
\email{annekatb@uci.edu}

\author{Tim M.P. Tait}
\email{ttait@uci.edu}
\affiliation{Department of Physics and Astronomy, University of California, Irvine, CA 92697 USA}
 
\author{Mauro Valli}%
 \email{mauro.valli@stonybrook.edu}
\affiliation{%
 C.N. Yang Institute for Theoretical Physics, Stony Brook University, Stony Brook, NY 11794,~USA
}%

\date{\today}

\begin{abstract}
The recent measurement of helium-4 from the near-infrared spectroscopy of extremely metal-poor galaxies (EMPGs) by the Subaru 
Survey may point to a new puzzle in the Early Universe.  We exploit this new helium measurement together with the percent-level determination of primordial deuterium, to 
assess indications for a non-vanishing lepton asymmetry during the Big Bang Nucleosynthesis (BBN) era, 
paying particular attention to the role of uncertainties in the nuclear reaction network. 
A cutting-edge Bayesian analysis focused on the role of the newly measured EMPGs, jointly with information from the Cosmic Microwave Background, suggests the existence of 
a nonzero lepton asymmetry at around the $2 \sigma$ level, 
providing a hint for cosmology beyond $\Lambda$CDM. 
We discuss conditions for a large total lepton asymmetry to be consistently realized in the Early Universe. 
\end{abstract}

\maketitle

% ##############################
\section{Introduction} 
\label{sec:intro}
% ##############################

Cosmological observations from the Early Universe provide an invaluable probe of Physics Beyond the Standard Model (BSM).   Observations of the
Cosmic Microwave Background (CMB), epitomized by the Planck mission~\cite{Planck:2018vyg} and further 
developed e.g. by the ACT~\cite{ACT:2020gnv} and SPT~\cite{SPT-3G:2021eoc} collaborations, 
paint a picture of a Universe dominated by non-baryonic dark energy
and dark matter, well-described by the $\Lambda$CDM model~\cite{Baumann:2015rya,Komatsu:2022nvu,Chang:2022tzj}.
Equipped with the CMB inference of the small cosmological baryonic abundance, $\Omega_{B} \sim 4\%$, the theory of
Big Bang Nucleosynthesis (BBN) within the Standard Model (SM) of Particle Physics is highly predictive, and confronted with
accurate measurements of primeval elements such as the mass density fraction of helium-4, ${\rm Y_P}$,
and the relative abundance of deuterium to hydrogen, ${\rm D/H}$, offers important constraints on New Physics (NP)~\cite{Sarkar:1995dd,Steigman:2007xt,Iocco:2008va,Pospelov:2010hj}
active during the first few minutes of the Universe~\cite{Serpico:2004nm,Boehm:2013jpa,Sabti:2019mhn,Giovanetti:2021izc}. 

At present, measurements of deuterium in quasar absorption spectra provide the best proxy for the determination of a primordial abundance. 
The most recent measurements from damped Lyman-$\alpha$ systems achieve better than $1\%$ precision~\cite{Cooke:2016rky,Riemer-Sorensen:2017pey,Cooke:2017cwo}, 
yielding a weighted average of ${\rm D/H} \times 10^5 = 2.547 \pm 0.025$~\cite{Zyla:2020zbs}.
This remarkable precision appears to be in tension with the SM at about the 2$\sigma$ level~\cite{Pitrou:2020etk}, 
although this remains under debate~\cite{Pisanti:2020efz,Yeh:2020mgl} in light of the uncertainties on the key nuclear reactions involved.
This highlights the primary importance to assess the impact of uncertainties in the nuclear network rates on the predictions from BBN~\cite{Pitrou:2021vqr}.
A notable recent advance in this direction is the improved determination of the $\textrm{D}(p,\gamma)^3\textrm{He}$ rate by the 
LUNA collaboration~\cite{Mossa:2020gjc}, which has an important impact on BBN constraints from primordial deuterium
on various NP scenarios~\cite{Sabti:2021reh}. 

\begin{figure}[!t!]
  \centering
  \includegraphics[width=0.49\textwidth]{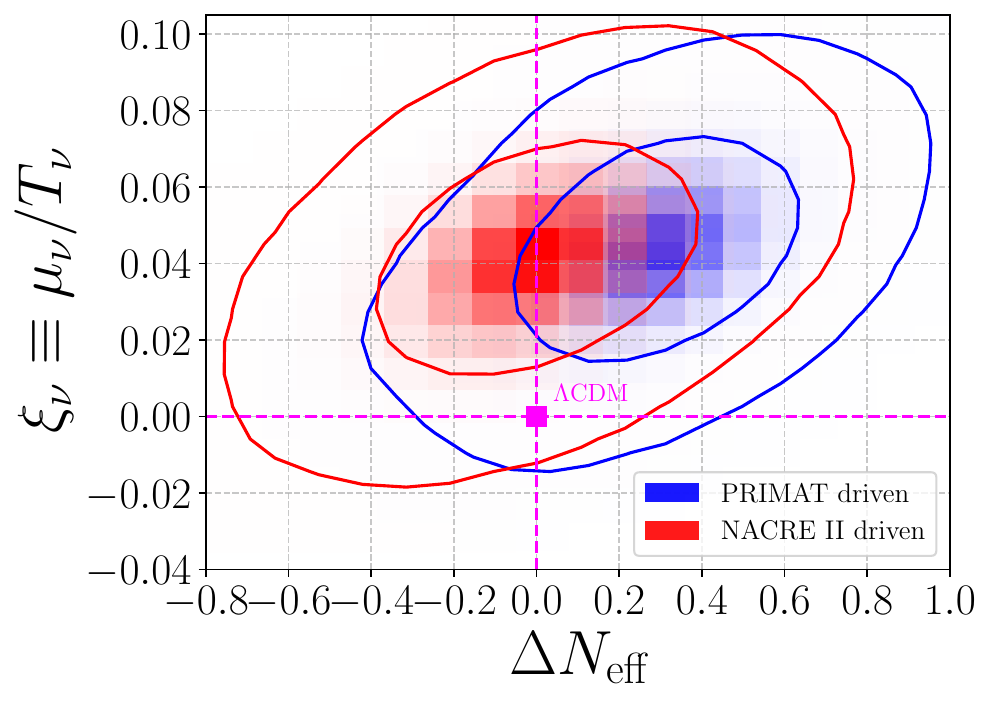}
  \caption{\it 68\% and 95\% two-dimensional probability distribution of the primordial chemical potential of neutrinos, $\mu_{\nu}$, 
  normalized to the neutrino temperature $T_{\nu}$, and the number of extra relativistic degrees of freedom in the Early Universe, $\Delta N_{\rm eff}$, 
  from a state-of-the-art analysis of BBN and CMB data. 
  The red and blue contours indicate the results for two different sets of nuclear uncertainties; magenta lines the $\Lambda$CDM prediction.}
  \label{fig:fig1}
\end{figure}

The recent near-infrared observation of 10 extremely metal-poor galaxies (EMPGs) by the Subaru Survey~\cite{Matsumoto:2022tlr} points
to even more puzzling mysteries. Spectroscopic observations of EMPGs provide a crucial input to the inference of ${\rm Y_P}$, because they
host the gas of nebulae resembling extraordinarily pristine environments which allow for a more accurate
extrapolation of the helium density to zero metallicity.
Combined with the pre-existing data from 3 EMPGs and 51 metal-poor galaxies~\cite{2020ApJ...896...77H} and measurements 
of the He $\lambda$10830 infrared emission line (relevant for parameter-degeneracy breaking~\cite{Aver:2015iza}), 5 (out of 10) Subaru EMPGs yield
a determination of primordial helium-4 of ${\rm Y_P} = 0.2370^{+0.0034}_{-0.0033}$, 
in sharp contrast with the PDG value ${\rm Y_P} = 0.245 \pm 0.003$~\cite{Zyla:2020zbs}, 
and well below the SM prediction~\cite{Pitrou:2020etk,Pisanti:2020efz,Yeh:2020mgl}, naively a 3$\sigma$-level discrepancy. 

Ref.~\cite{Matsumoto:2022tlr} took the first steps toward an interpretation of this \textit{helium anomaly} in terms of a BSM fit where the standard theory has 
been extended by extra-relativistic degrees of freedom, $\Delta N_{\rm eff}$, as well as a nonzero electron neutrino asymmetry, 
$\xi_{\nu_e}$, while simply anchoring $\Omega_{B}h^2$ to the most precise determination derived by Planck~\cite{Planck:2018vyg}. 

In this \textit{Letter} we revise the inference of a 
lepton asymmetry $\xi_{\nu}$ in the Early Universe, as well as on $\Delta N_{\rm eff}$ (defined at the last scattering), paying attention to the details of a joint likelihood analysis of 
BBN and CMB data as recently carefully formulated in~\cite{Sabti:2019mhn,Giovanetti:2021izc}. 
Our key result is given in Figure~\ref{fig:fig1}~\footnote{While the methodology developed in this work stands out as a robust recipe for an advanced statistical analysis of Early Universe data, the measurement published in~\cite{Matsumoto:2022tlr}, focus of the present study, strongly depends on the emission-line data modeling. Further scrutiny on the corresponding systematics of the measurement is warranted for the future.}.
We perform a Bayesian analysis taking into account the theory uncertainties pivotal for unbiased conclusions
based on the use of the new public code for state-of-the-art investigations of (BSM) physics in the 
Early Universe -- \texttt{PRyMordial} -- presented in a companion paper~\cite{PRyM}.

%Figure~\ref{fig:fig1} shows our key result: a nonzero asymmetry in the lepton sector is favored by current data at $90\%$ probability, 
%irrespective of nuclear modeling. At the same time, with somewhat more dependence on the treatment of the BBN nuclear network rates, 
%the data may provide some hint for new sources of relativistic degrees of freedom. 
%In the following, we detail the outcome of the present analysis and comment on the possible NP implications.

% ##############################
\section{Primordial Lepton Asymmetries} 
\label{sec:inf}
% #############################

Electric charge neutrality of the Early Universe does not allow for a large primordial asymmetry in the charged lepton sector, which is constrained to
be (at most) of the order of the baryon-to-photon ratio $\eta_{B}\equiv n_{B}/n_{\gamma}\sim \mathcal{O}(10^{-10})$~\cite{Kolb:1990vq,Rubakov:2017xzr}. 
Nevertheless, a large cosmic asymmetry can be hidden in the neutrino sector~\cite{Simha:2008mt}: 
\begin{equation}
    \label{eq:etaL}
    \eta_L \equiv \frac{1}{n_{\gamma}}\sum_{i=e,\mu,\tau}(n_{\nu_i}-n_{\bar{\nu}_i}) \simeq \frac{\pi^2}{33 \zeta(3)} (\xi_{\nu_e}+\xi_{\nu_\mu}+\xi_{\nu_\tau}) \ ,
\end{equation}
where $n_{\gamma}$ is the photon number density, $n_{\nu_{i}}$ the flavor $i$ neutrino density, and 
$\xi_{\nu_i} \equiv \mu_{\nu_i}/T_{\nu_i}$  are the \textit{degeneracy parameters} defined as
the chemical potential for each neutrino normalized to its temperature, which encode the relevant lepton asymmetries today. 
Eq.~\eqref{eq:etaL} assumes $T_{\nu_{i}}/T_{\gamma} = (4/11)^{1/3}$, which is a good approximation given the modest impact of non-instantaneous neutrino decoupling 
and tiny departures from the Fermi-Dirac distributions in relativistic freeze-out~\cite{Akita:2020szl,Froustey:2020mcq,Bennett:2020zkv}.
It is further relevant that nonzero neutrino chemical potentials play a marginal role in SM neutrino decoupling~\cite{Grohs:2015tfy,EscuderoAbenza:2020cmq}. 

Eq.~\eqref{eq:etaL} further implements the condition $|\xi_{\nu_i}| < 1$, as $\mathcal{O}(1)$ degeneracy parameters 
were probed by early-stage CMB observations about two decades ago~\cite{Kinney:1999pd,Lesgourgues:1999wu},
and now are robustly~\cite{Oldengott:2017tzj,Bonilla:2018nau,Kumar:2022vee} ruled out (irrespective of the lepton flavor~\cite{Abazajian:2002qx,Castorina:2012md}). 
In fact, a nonzero chemical potential for the $i$-flavored neutrino would yield a contribution to the total radiation density (relative to photons) of
\begin{equation}
    \label{eq:deltarhorad}
    \frac{\Delta \rho_{\rm rad}}{\rho_{\gamma}} \simeq \frac{15}{4\pi^2}\left(\frac{4}{11} \right)^{4/3} \xi_{\nu_{i}}^2 \ ,
\end{equation}
and would increase the expansion rate of the Universe, resulting in a positive shift of $N_{\rm eff}$ that
would delay the time of matter-radiation equality which is tightly constrained by the CMB acoustic peaks. 

From the Planck constraint on $N_{\rm eff}$ adopting the likelihood analysis including TTTEEE and low-$\ell$ measurements, as well as
baryonic acoustic oscillations (BAO) and lensing data, and assuming a flat prior on $\rm Y_{\rm P}$, one may derive a simple upper-bound on the 
degeneracy parameters purely driven by the CMB.
In particular, for $N_{\rm eff} = 2.97 \pm 0.29$ (68\% probability interval)~\cite{Planck:2018vyg,Planck18res}, 
considering the SM prediction $N_{\rm eff} = 3.044$ (known better than the per-mille level)~\cite{Akita:2020szl,Froustey:2020mcq,Bennett:2020zkv}, 
the 1$\sigma$ upper-bound is:
\begin{equation}
\label{eq:xiallflav}
\xi_{\nu_{e}}^2 + \xi_{\nu_{\mu}}^2 + \xi_{\nu_{\tau}}^2 \lesssim 0.5 \ ,
\end{equation}
implying the conservative constraint $|\xi_{\nu_{i}}| \lesssim 0.71$, valid for each flavor individually (see also~Ref.~\cite{Barenboim:2016lxv}). 
Since the onset of neutrino oscillations is expected to occur around $T_{\nu} \sim 10$~MeV, 
flavor equilibration in the muon-tau sector is predicted to be complete by the time of neutrino decoupling ($T_{\nu} \sim 2$~MeV)~\cite{Dolgov:2002ab,Pastor:2008ti},
and the conservative CMB bound of Eq.~\eqref{eq:xiallflav} becomes slightly tighter for the 2$^{\rm nd}$ and 3$^{\rm rd}$ generation $\nu$ asymmetries:
\begin{equation}
\label{eq:ximutau}
|\xi_{\nu_{\mu,\tau}}| \lesssim 0.5 \ .
\end{equation}

BBN can place stronger constraints on the electron-neutrino asymmetry (by about an order of magnitude~\cite{Simha:2008mt,Fields:2019pfx}) largely
because an electron-neutrino asymmetry at the time of BBN affects the $\beta$ equilibrium of weak interactions controlling the neutron-proton conversion~\cite{Kohri1997}.
A positive (negative) value of $\xi_{\nu_e}$ acts through the equilibrium
reactions $n \, \nu_{e} \leftrightarrow p \, e^-$, $p \, \bar{\nu}_{e} \leftrightarrow n \, e^+$ and neutron decay to reduce (enhance) the neutron-to-proton ratio:
\begin{equation}
    \label{eq:ntopratio}
    (n_{n}/n_{p})\big|_{\rm eq.} \simeq \exp\left(-\mathcal{Q}/T_{\gamma}-\xi_{\nu_e}\right) \ ,
\end{equation}
where $\mathcal{Q} \equiv m_{n}-m_{p} = 1.293$~MeV is the neutron-proton mass difference. 
While light primordial abundances like deuterium are particularly sensitive to $\Omega_{B}$, from Eq.~\eqref{eq:ntopratio} it follows that helium-4 -- which depends 
crucially on the amount of neutrons at the time where deuterium is no longer photo-dissociated~\cite{Kolb:1990vq,Rubakov:2017xzr} -- 
can be regarded as a sensitive \textit{primordial leptometer}.

In Ref.~\cite{Pitrou:2018cgg} a combined analysis of the helium and deuterium PDG
values together with a Gaussian prior on $\Omega_{B}h^2$ from the CMB yields the precise determination: $\xi_{\nu} = 0.001 \pm 0.016$, consistent with zero.
By assuming full equilibration of lepton flavor asymmetries due to neutrino oscillations, this inference is more stringent than the bound outlined in Eq.~\eqref{eq:ximutau}. 
Nevertheless, a recent state-of-the-art investigation in Ref.~\cite{Froustey:2021azz} indicates that the degree to which full flavor equilibration is realized during 
the BBN era sensitively depends on the PMNS mixing angle $\theta_{13}$ and on the initially generated values of the degeneracy parameters.
In the following, we revisit the determination of $\xi_{\nu}$ in light of the newly measured helium-4 mass fraction from EMPGs as reported in Ref.~\cite{Matsumoto:2022tlr}. 
While $\xi_{\nu} = \xi_{\nu_{e,\mu,\tau}}$ may be achieved in the Early Universe, 
one should bear in mind that the conservative interpretation of our main finding in Fig.~\ref{fig:fig1} applies only to 
$\xi_{\nu} = \xi_{\nu_{e}}$, i.e. the primordial electron-neutrino asymmetry probed by BBN via $\beta$ equilibrium, Eq.~\eqref{eq:ntopratio}.

% ##############################
\section{Methodology}
\label{sec:met}
% ##############################

\begin{figure*}[!t!]
  \centering
  \includegraphics[width=\textwidth]{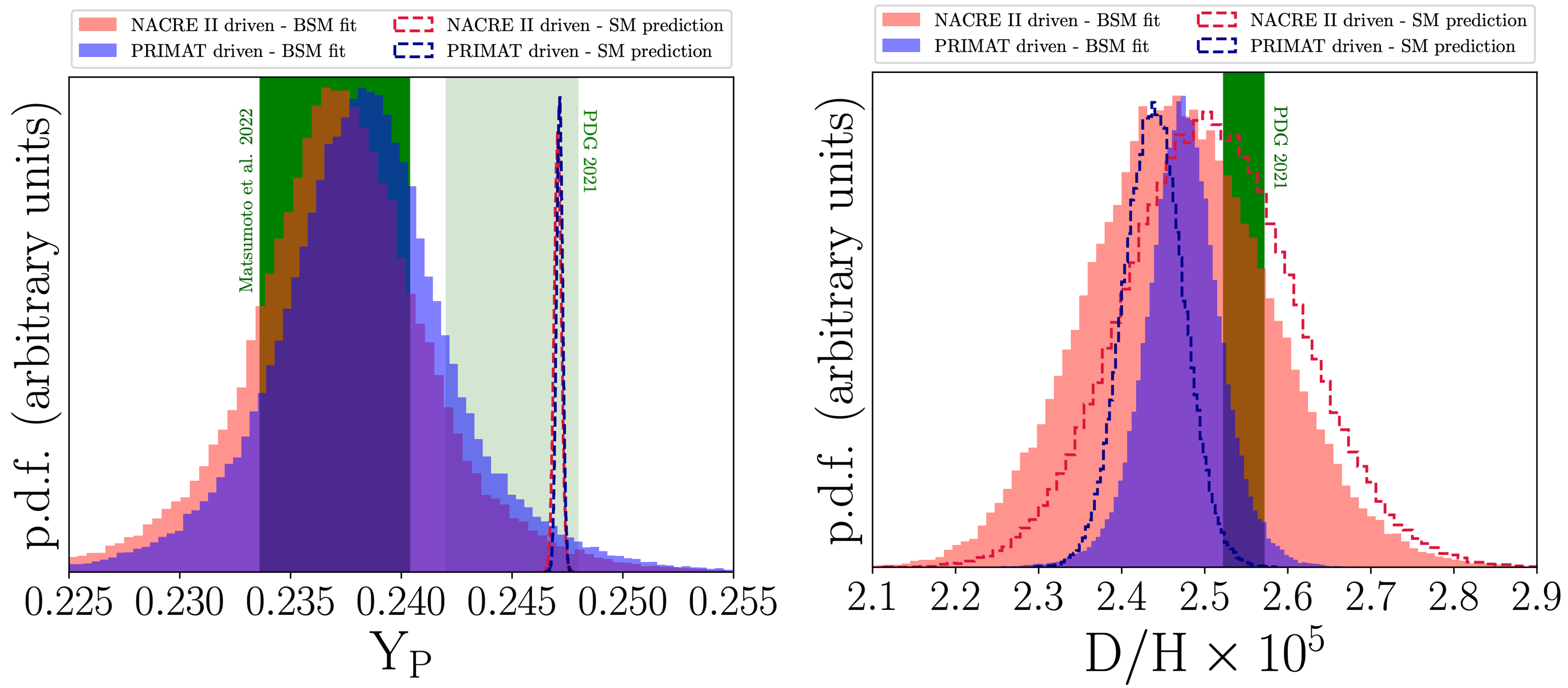}
  \caption{\it Probability density function (p.d.f.) for the primordial light elements analyzed in this study. In the left panel, the p.d.f. for helium-4, $Y_{\rm P}$, as precisely predicted in the SM according to two different set of nuclear uncertainties and adopting the determination of the cosmological baryon-to-photon ratio from the fit of CMB data within $\Lambda$CDM (color code similar to Figure~\ref{fig:fig1}). In the same panel, the outcome from the joint fit to BBN and CMB likelihoods in the BSM scenario where $\xi_{\nu}$ and $\Delta N_{\rm eff}$ are consistently allowed to differ from their $\Lambda$CDM limit. In the right panel, the same set of p.d.f.s is shown for the deuterium. In both panels, vertical dark green bands correspond to the 1$\sigma$ interval for the BBN measurements employed in the analysis. In the left one, the PDG 2021 recommended value for helium-4 is also reported, in agreement with the SM prediction.}
  \label{fig:fig2}
\end{figure*}

Our computation of BBN abundances via the \texttt{PRyMordial}~\cite{PRyM} code, proceeds through 3 main steps:
\begin{enumerate}
    \item[\textit{1)}] Solving for the thermal background; % comprising the electron-photon plasma and neutrino species, to pin down the scale factor of the universe as a function of $T_{\gamma}$ and time, as well as the relation $T_{\nu}(T_{\gamma})$;
    \item[\textit{2)}] Computing neutron-proton conversion;% via the weak effective theory, determining the initial conditions for the BBN nuclear network;
    \item[\textit{3)}] Evaluating the final primordial abundances. % at the end of BBN. according to a set of reactions that are key for precise helium-4 and deuterium predictions.
\end{enumerate}

For \textit{1)} we base our computation on the approach proposed in~\cite{Escudero:2018mvt} and further developed in~\cite{EscuderoAbenza:2020cmq} (see also~\cite{Chu:2022xuh}). It consists in solving the Boltzmann equations for the electron-photon plasma and neutrinos assuming a thermal distribution for the species, including NLO QED corrections for the plasma~\cite{Bennett:2019ewm} as well as non-instantaneous decoupling effects for the neutrino sector~\cite{EscuderoAbenza:2020cmq}. For our purposes, it suffices to describe the neutrino sector by a common temperature $T_{\nu}$, yielding the SM prediction $N_{\rm eff} = 3.045$, differing from the most refined prediction in~\cite{Akita:2020szl,Froustey:2020mcq,Bennett:2020zkv} only at the per-mille level, well within current and future observational  sensitivity~\cite{SimonsObservatory:2018koc,CMB-S4:2022ght,CMB-HD:2022bsz}.  A nonzero chemical potential for neutrinos would influence our analysis of the thermal background via Eq.~\eqref{eq:deltarhorad}. If full neutrino equilibration is achieved, we find a-posteriori a contribution to the radiation density that would be totally negligible. Nevertheless, in our BSM analysis we also account for the possibility of  a nonzero lepton asymmetry $ |\xi_{\nu_{\mu,\tau}}| \gg |\xi_{\nu_{e}}|$ by varying $\Delta N_{\rm eff}$ (a valid interpretation of $\Delta N_{\rm eff}$ in the scenario where such a shift is mainly driven by a non-vanishing muon-tau chemical potential). Note that from our bound in Eq.~\eqref{eq:ximutau}, a muon-tau neutrino asymmetry can induce a maximal shift $ \Delta N_{\rm eff} \sim 0.1$.

Moving to \textit{2)}, we compute $n \leftrightarrow p$ matrix elements beyond the Born approximation~\cite{Sirlin:1967zza}, namely including isospin-breaking contributions like finite-mass~\cite{Brown:2000cp} and QED~\cite{Czarnecki:2004cw,Ivanov:2017fra} corrections, as well as finite-temperature effects~\cite{Lopez:1997ki}, following the implementation carried out in~\cite{Pitrou:2018cgg}. Most importantly, we evaluate weak-interaction rates integrating over nucleon thermal distributions with chemical potential $\mu_{\mathcal{Q}} \equiv \mu_{n}-\mu_{p} = - \mu_{\nu_e} \neq 0$. 

Finally, regarding \textit{3)}, we proceed evolving the abundances according to the network of thermonuclear reactions comprising the main processes listed in~\textbf{Table~1} of Ref.~\cite{Iliadis:2020jtc} (plus $^3\textrm{He}(p,\gamma)^4\textrm{He}$, taken from~\cite{Serpico:2004gx}), yielding state-of-the-art predictions for $Y_{\rm P}$ and ${\rm D/H}$. In particular, for the radiative neutron capture rate we adopt the MCMC result of Ref.~\cite{Ando:2005cz}, while in the treatment of the other 10 key reactions we distinguish two approaches:
\begin{itemize}
    \item[$\circ$] \textcolor{newblue}{PRIMAT driven:} Nuclear rates are implemented according to the statistical determination of Refs~\cite{Descouvemont:2004cw,Iliadis:2016vkw,InestaGomez:2017eya,deSouza:2018gdx,deSouza:2019pmr,Moscoso:2021xog}, i.e. following theoretical ab-initio energy modeling tuned to datasets for which an estimate of systematic errors is available~\cite{Pitrou:2018cgg,Pitrou:2020etk,Pitrou:2021vqr}.
    \item[$\circ$] \textcolor{newred}{NACRE~II driven:} Nuclear rates are interpolated from the updated NACRE compilation~\cite{Xu:2013fha}, comprising charged-particle-induced reactions~\footnote{In particular, for the key reactions $\textrm{D}(d,n)^3\textrm{He}$ and $\textrm{D}(d,p)^3\textrm{H}$ we either interpolate rates and corresponding uncertainties from the NACRE~II numerical database or from the outcome of the Bayesian analysis of Ref.~\cite{Iliadis:2016vkw}.}; for $\textrm{D}(p,\gamma)^3\textrm{He}$ we use the LUNA result reported in~\cite{Mossa:2020gjc}; for $^7\textrm{Be}(n,p)^7\textrm{Li}$ we adopt the baseline of Ref.~\cite{Fields:2019pfx}.
\end{itemize}

We perform a Bayesian analysis of Early Universe data constructing the cosmological test statistic:
\begin{equation}
    \label{eq:TScosmo}
    \textrm{TS}_{\rm cosmo} \equiv -2 (\log \mathcal{L}_{\rm CMB} + \log \mathcal{L}_{\rm BBN}) \ ;
\end{equation}
the CMB likelihood explicitly reads: 
\begin{equation}
    \label{eq:logCMB}
    \log \mathcal{L}_{\rm CMB} = -\frac{1}{2}  \Delta \vec{v} ^{\,T} \, \mathcal{C}^{-1}_{\rm CMB} \, \Delta \vec{v} \ ,
\end{equation}
with $\Delta \vec{v} \equiv \vec{v}^{\,\rm th}-\vec{v}$, $ \vec{v} = ({\rm Y_P}, \Omega_{B}h^{2}, N_{\rm eff})^T \ $,
using mean and standard-deviation values from the TTTEEE + low-$\ell$ + BAO + lensing Planck run varying also ${\rm Y_{P}}$, $N_{\rm eff}$~\cite{Planck:2018vyg,Planck18res} and also retrieving correlations in $\mathcal{C}_{\rm CMB}$ from~\cite{Planck18download}.  The BBN likelihood of our study corresponds to:
\begin{equation}
    \label{eq:logBBN}
    \log \mathcal{L}_{\rm BBN} = -\frac{1}{2}  \sum_{X} \left(\frac{X^{\rm th}-X}{\sigma_{X}}\right)^2 \ ,
\end{equation}
where $X = \{{\rm Y_P}, {\rm D/H}$\}, and we use the measurements: ${\rm Y_P} = 0.2370(34)$~\cite{Matsumoto:2022tlr}, ${\rm D/H} = 0.00002547(25)$~\cite{Zyla:2020zbs}.

The parameters we infer are varied according to uniform priors: $-2 \leq \Delta N_{\rm eff} \leq 2$, $ \ -0.2 \leq \xi_{\nu} \leq 0.2$,  $ 1 \leq (\eta_{B} \times 10^{10}) \leq 10$ (using $\eta_{B} \times 10^{10}\simeq  273.748\, \, \Omega_{B}h^2 $). We marginalize over the neutron lifetime and the adopted nuclear uncertainties. From the PDG analysis~\cite{Zyla:2020zbs} we assign  the Gaussian prior: $\tau_{n} = (879.4 \pm 0.6$)~s to the neutron lifetime. For the uncertainties in the nuclear rates, we assign log-normal distributions following the method detailed in Ref.~\cite{Coc:2014oia}, varying a total of 12 additional nuisance parameters.

We  perform an MCMC analysis via the \texttt{emcee}~\cite{emcee2013} package, using 60 walkers with 2100 steps each, discarding the first 700 steps of each walker as burn-in.
From the best-fit values minimizing $\textrm{TS}_{\rm cosmo}$ we also compute for each scenario the Information Criterion~\cite{AkaikeIC,2013arXiv1307.5928G} $IC \equiv -2 \log \hat{\mathcal{L}}_{\rm BBN} + 2(k-1)$, $k$ being the number of BSM parameters and accounting for the CMB information as an extra constraint in the fit. Then, we evaluate the $IC$ difference with respect to the SM prediction of the primordial light abundances within a given approach: $\Delta IC \sim \mathcal{O}(1)$ ($\sim \mathcal{O}(10)$) provides positive (strong) support in favor of NP beyond $\Lambda$CDM according to the canonical scales of evidence~\cite{BayesFactors}.

% ##############################
\section{Results}
\label{sec:res}
% ##############################

\begin{table*}[!ht!]
\centering
\renewcommand{\arraystretch}{1.5}
{\footnotesize
\centering
\begin{tabular}{|c||c||c|c||c|c|c||c|}
\hline
\textbf{Scenario} & \textbf{Approach} & \boldmath$ \ \rm Y_P \ $ & \boldmath$ \ {\rm D/H} \times 10^5 \ $ & \boldmath$ \ \Delta N_{\rm eff} \ $
& \boldmath$ \ \xi_{\nu} \ $ & \boldmath$ \ \eta_{B} \times 10^{10} \ $ & \boldmath$ \ \Delta IC \ $ \\
\hline
\multirow{2}{*}{\textit{SM prediction}}
& \textcolor{newblue}{ PRIMAT driven } &
0.24715(14)  & 2.439(36) & --  & --  & 6.137(38)  & --  \\ 
 & \textcolor{newred}{ NACRE II driven } &
0.24706(16)  & 2.51(10) & --  & --  & 6.137(38)  & --  \\ 
\hline
\multirow{2}{*}{\textit{$\Delta  N_{\rm eff}$ BSM fit}}
& \textcolor{newblue}{ PRIMAT driven } &
 0.2472(13)  &  2.472(45) & 0.02(20)  & --  & 6.091(66)  & \textcolor{newblue}{2}  \\ 
 & \textcolor{newred}{ NACRE II driven } &
0.2455(15) & 2.46(11) & -0.26(23)  & --  & 6.093(67)  & \textcolor{newred}{0}  \\ 
\hline
%
%\multirow{2}{*}{\textit{$\xi_{\nu}$ BSM fit}}
%& \textcolor{newblue}{ PRIMAT driven } &
%2.423(39)  & 2.47(10) & --  & 0.020(16)  & 6.05(13)  & \textcolor{newblue}{xxx}  \\ 
% & \textcolor{newred}{ NACRE II driven } &
%2.383(39)  & 2.48(15) & --  & 0.037(16)  & 6.11(14)  & \textcolor{newred}{xxx}  \\ 
 %
%\hline
%
\multirow{2}{*}{\textit{ ($ \Delta  N_{\rm eff},\xi_{\nu}$) BSM fit }}
& \textcolor{newblue}{ PRIMAT driven } &
0.2383(42)  & 2.474(46) & 0.29(23)  & 0.044(20)  & 6.119(65)  & \textcolor{newblue}{8}  \\ 
 & \textcolor{newred}{ NACRE II driven } &
0.2372(43)  & 2.47(11) & 0.00(23)  & 0.041(21)  & 6.114(68)  & \textcolor{newred}{5}  \\  
\hline
\end{tabular}
}
\caption{\it 68\% probability interval for the posterior distribution of the main observables and parameters in the scenarios considered in this work. For the BSM fits, improvement with respect to the SM is given by $\Delta IC>0$, see text for more details.}
\label{tab:res}
\end{table*}

In Figure~\ref{fig:fig1} we report the main result of our study: the 68\%  and 95\% probability region for the primordial lepton asymmetry $\xi_{\nu}$ and the extra-relativistic degrees of freedom~$\Delta N_{\rm eff}$ as determined by TS$_{\rm cosmo}$, Eq.~\eqref{eq:TScosmo}, corresponding to the two approaches to thermonuclear rates described in the previous section. From the $\Lambda$CDM limit highlighted in the same figure, we can conclude that a BSM fit to a dataset that includes the newly measured EMPGS by Subaru~\cite{Matsumoto:2022tlr} favors at present a non-vanishing asymmetry in the neutrino sector.

In Figure~\ref{fig:fig1} we also observe that, dependent on the approach to nuclear uncertainties, a shift of $N_{\rm eff}$ of $\mathcal{O}(1)$ can be simultaneously favored by current cosmological data. Note that the size of the shift in the number of relativistic degrees of freedom can be interpreted within the 68\% probability region as the result of a large neutrino asymmetry in the muon-tau sector in case flavor equilibration has not been fully realized.

To further investigate the different outcome from each approach, we show in Figure~\ref{fig:fig2} the posterior probability density function (p.d.f.) for the BBN observables  ${\rm Y_{P}}$ and ${\rm D/H}$.  We report both the result from the BSM fit varying both $\xi_{\nu}$ and $N_{\rm eff}$ as well as the one from the SM prediction, obtained fixing the BSM parameters to 0 and replacing the CMB likelihood with the Gaussian prior: $\Omega_{B} h^2 = 0.02242 \pm 0.00014$, from the $\Lambda$CDM Planck analysis (TTTEEE + low-$\ell$ + BAO + lensing)~\cite{Planck:2018vyg,Planck18res}. In the same figure, we also highlight with vertical dark green bands the measurements adopted in our BBN analysis via Eq.~\eqref{eq:logBBN}, and report the PDG~2021 value ${\rm Y_P} = 0.245(3)$~\cite{Zyla:2020zbs}, in optimal agreement with the analysis of Ref.~\cite{Xu:2013fha} that comprises the set studied also in~\cite{Matsumoto:2022tlr} without the new EMPGs from Subaru.

Figure~\ref{fig:fig2} neatly highlights two tensions in the limit where BSM physics is not accounted for:
\begin{itemize}
    \item A discrepancy at the 3$\sigma$ level between the SM prediction of ${\rm Y_{P}}$ and the newly inferred helium-4 mass-fraction value, regardless of the approach taken for the thermonuclear reactions; the tension is fully driven by the new measurement delivered by Ref.~\cite{Matsumoto:2022tlr}, while the overall significance also depends on the precision obtained for the inference of the cosmological baryon abundance within $\Lambda$CDM;
    \item A tension of about 2$\sigma$ significance between the SM prediction of D/H and the PDG 2021 recommended measurement~\cite{Zyla:2020zbs} when the PRIMAT driven approach is taken for the analysis of the key thermonuclear reactions involved, in line with recent discussions in the literature~\cite{Pitrou:2021vqr}.
\end{itemize}

From Figure~\ref{fig:fig2}, it is clear that a shift of $\Delta N_{\rm eff}$ is required together with $\xi_{\nu} \neq 0$ only when the PRIMAT driven approach is considered, in order to address the discrepancy consequently present in the fit in relation to the observed primordial deuterium abundance. 
In the same figure it is also evident how the PDG 2021 recommended measurement of the helium-4 mass fraction is in perfect agreement with the SM prediction, and our inference for a nonzero degeneracy parameter $\xi_{\nu}$ is the consequence of adopting the new ${\rm Y_{P}}$ measurement~\cite{Matsumoto:2022tlr}.

We report in Table~\ref{tab:res} the $68\%$ probability interval for the scenarios discussed so far as well as the one for the BSM fit where only $\Delta N_{\rm eff}$ is considered. Looking at the $\Delta IC$ values, we conclude that a joint analysis of BBN + CMB data provides mild to strong evidence for a scenario with non-vanishing lepton asymmetry. Moreover, within the NACRE II approach no notable support from data is found for the presence of extra relativistic degrees of freedom in the Early Universe, 
whereas a scenario where only $\Delta N_{\rm eff}$ is varied may be slightly preferred by data over the SM in the case of the PRIMAT driven approach, partially ameliorating a potential \textit{deuterium anomaly}.

% ##############################
\section{Discussion and outlook}
\label{sec:disc}
% ###############################

%We find that the new measurement of primeval helium including 10 additional EMPGs by the Subaru Survey leads to a tantalizing hint for a large total lepton asymmetry of primordial origin. 
Our study based on the addition of the newly observed EMPGs~\cite{Matsumoto:2022tlr} to the original sample of ref.~\cite{2020ApJ...896...77H} suggests that today's total lepton asymmetry, Eq.~\eqref{eq:etaL}, is large, $\eta_{L} \gg \eta_{B}$, ranging from $\sim 10^{-2}$ to $\sim 1/4$, depending on the details of the neutrino sector.
%oscillations in the Early Universe as well as on the initial value for the neutrino degeneracy parameters. Such a finding would require a cosmology dramatically different from the one inferred extrapolating the SM, and thus hints toward BSM physics.  

%While it is possible to imagine many types of NP which could generate such a situation, 
There are common factors that any successful explanation of a large $\eta_{L}$ must share.
At temperatures above the scale of electroweak symmetry restoration, electroweak sphalerons equilibrate
$B + L$ such that the final total lepton and baryon asymmetries differ by a ${\cal O}(1)$ factor \cite{Khlebnikov:1988sr,Harvey:1990qw}.  Thus, for a difference of orders of magnitude between $\eta_{L}$ and $\eta_{B}$
to persist, it must either be generated after the sphalerons become inactive (in the SM, at the electroweak phase transition around temperatures of order 100 GeV) or
the individual flavor asymmetries must be distributed such that the net $L$ is much smaller than the individual asymmetries \cite{March-Russell:1999hpw}. The latter scenario would point to
flavor-dependent NP in the lepton sector, with possible interesting implications for the smallness of $\eta_B$ as well, see for instance~\cite{Dreiner:1992vm,March-Russell:1999hpw,Shu:2006mm,Gu:2010dg}.  

Because equilibration of neutrino species
depends both on imprecisely determined mixing parameters and the assumed initial asymmetry in each flavor~\cite{Froustey:2021azz}, mapping the inferred neutrino asymmetries during BBN into the space of consistent initial conditions at some earlier time is an interesting inverse problem; it requires also assumptions on the interpretation of the inference carried out here for $\Delta N_{\rm eff}$, and it is beyond the scope of this work.  Several examples of theories capable of generating a sufficiently large and persistent lepton-flavored neutrino asymmetry
via variations of the Affleck-Dine mechanism \cite{Affleck:1984fy}
exist in the literature~\cite{Casas:1997gx,McDonald:1999in,March-Russell:1999hpw,Kawasaki:2022hvx}.

%Looking forward, our study highlights the combined power of BBN and CMB data to provide a key glimpse of an early epoch of the Universe.  Making the most of its opportunity requires more work, to improve
%the treatment of nuclear physics inputs and measurements of the primordial abundances.  If the current indications of a large total lepton asymmetry $\eta_{L}$ survives further scrutiny, it provides
%important clues towards our understanding of the Universe in its early stages, offering at the same time a quite privileged view on BSM physics.

~
\begin{acknowledgments}
We thank Federico Bianchini, Kevork Abazajian, Manoj Kaplinghat, Rouven Essig and Peizhi Du for discussion.
The work of M.V. is supported by the Simons Foundation under the Simons Bridge for Postdoctoral Fellowships at SCGP and YITP, award number 815892.
The work of TMPT is supported by the US National Science Foundation under grant no.~PHY-1915005.
\end{acknowledgments}

\bibliography{biblio} % Produces the bibliography via BibTeX.

\begin{thebibliography}{90}%
\makeatletter
\providecommand \@ifxundefined [1]{%
 \@ifx{#1\undefined}
}%
\providecommand \@ifnum [1]{%
 \ifnum #1\expandafter \@firstoftwo
 \else \expandafter \@secondoftwo
 \fi
}%
\providecommand \@ifx [1]{%
 \ifx #1\expandafter \@firstoftwo
 \else \expandafter \@secondoftwo
 \fi
}%
\providecommand \natexlab [1]{#1}%
\providecommand \enquote  [1]{``#1''}%
\providecommand \bibnamefont  [1]{#1}%
\providecommand \bibfnamefont [1]{#1}%
\providecommand \citenamefont [1]{#1}%
\providecommand \href@noop [0]{\@secondoftwo}%
\providecommand \href [0]{\begingroup \@sanitize@url \@href}%
\providecommand \@href[1]{\@@startlink{#1}\@@href}%
\providecommand \@@href[1]{\endgroup#1\@@endlink}%
\providecommand \@sanitize@url [0]{\catcode `\\12\catcode `\$12\catcode
  `\&12\catcode `\#12\catcode `\^12\catcode `\_12\catcode `\%12\relax}%
\providecommand \@@startlink[1]{}%
\providecommand \@@endlink[0]{}%
\providecommand \url  [0]{\begingroup\@sanitize@url \@url }%
\providecommand \@url [1]{\endgroup\@href {#1}{\urlprefix }}%
\providecommand \urlprefix  [0]{URL }%
\providecommand \Eprint [0]{\href }%
\providecommand \doibase [0]{http://dx.doi.org/}%
\providecommand \selectlanguage [0]{\@gobble}%
\providecommand \bibinfo  [0]{\@secondoftwo}%
\providecommand \bibfield  [0]{\@secondoftwo}%
\providecommand \translation [1]{[#1]}%
\providecommand \BibitemOpen [0]{}%
\providecommand \bibitemStop [0]{}%
\providecommand \bibitemNoStop [0]{.\EOS\space}%
\providecommand \EOS [0]{\spacefactor3000\relax}%
\providecommand \BibitemShut  [1]{\csname bibitem#1\endcsname}%
\let\auto@bib@innerbib\@empty
%</preamble>
\bibitem [{\citenamefont {Aghanim}\ \emph {et~al.}(2020)\citenamefont {Aghanim}
  \emph {et~al.}}]{Planck:2018vyg}%
  \BibitemOpen
  \bibfield  {author} {\bibinfo {author} {\bibfnamefont {N.}~\bibnamefont
  {Aghanim}} \emph {et~al.} (\bibinfo {collaboration} {Planck}),\ }\href
  {\doibase 10.1051/0004-6361/201833910} {\bibfield  {journal} {\bibinfo
  {journal} {Astron. Astrophys.}\ }\textbf {\bibinfo {volume} {641}},\ \bibinfo
  {pages} {A6} (\bibinfo {year} {2020})},\ \bibinfo {note} {[Erratum:
  Astron.Astrophys. 652, C4 (2021)]},\ \Eprint
  {http://arxiv.org/abs/1807.06209} {arXiv:1807.06209 [astro-ph.CO]}
  \BibitemShut {NoStop}%
\bibitem [{\citenamefont {Aiola}\ \emph {et~al.}(2020)\citenamefont {Aiola}
  \emph {et~al.}}]{ACT:2020gnv}%
  \BibitemOpen
  \bibfield  {author} {\bibinfo {author} {\bibfnamefont {S.}~\bibnamefont
  {Aiola}} \emph {et~al.} (\bibinfo {collaboration} {ACT}),\ }\href {\doibase
  10.1088/1475-7516/2020/12/047} {\bibfield  {journal} {\bibinfo  {journal}
  {JCAP}\ }\textbf {\bibinfo {volume} {12}},\ \bibinfo {pages} {047} (\bibinfo
  {year} {2020})},\ \Eprint {http://arxiv.org/abs/2007.07288} {arXiv:2007.07288
  [astro-ph.CO]} \BibitemShut {NoStop}%
\bibitem [{\citenamefont {Dutcher}\ \emph {et~al.}(2021)\citenamefont {Dutcher}
  \emph {et~al.}}]{SPT-3G:2021eoc}%
  \BibitemOpen
  \bibfield  {author} {\bibinfo {author} {\bibfnamefont {D.}~\bibnamefont
  {Dutcher}} \emph {et~al.} (\bibinfo {collaboration} {SPT-3G}),\ }\href
  {\doibase 10.1103/PhysRevD.104.022003} {\bibfield  {journal} {\bibinfo
  {journal} {Phys. Rev. D}\ }\textbf {\bibinfo {volume} {104}},\ \bibinfo
  {pages} {022003} (\bibinfo {year} {2021})},\ \Eprint
  {http://arxiv.org/abs/2101.01684} {arXiv:2101.01684 [astro-ph.CO]}
  \BibitemShut {NoStop}%
\bibitem [{\citenamefont {Baumann}\ \emph {et~al.}(2016)\citenamefont
  {Baumann}, \citenamefont {Green}, \citenamefont {Meyers},\ and\ \citenamefont
  {Wallisch}}]{Baumann:2015rya}%
  \BibitemOpen
  \bibfield  {author} {\bibinfo {author} {\bibfnamefont {D.}~\bibnamefont
  {Baumann}}, \bibinfo {author} {\bibfnamefont {D.}~\bibnamefont {Green}},
  \bibinfo {author} {\bibfnamefont {J.}~\bibnamefont {Meyers}}, \ and\ \bibinfo
  {author} {\bibfnamefont {B.}~\bibnamefont {Wallisch}},\ }\href {\doibase
  10.1088/1475-7516/2016/01/007} {\bibfield  {journal} {\bibinfo  {journal}
  {JCAP}\ }\textbf {\bibinfo {volume} {01}},\ \bibinfo {pages} {007} (\bibinfo
  {year} {2016})},\ \Eprint {http://arxiv.org/abs/1508.06342} {arXiv:1508.06342
  [astro-ph.CO]} \BibitemShut {NoStop}%
\bibitem [{\citenamefont {Komatsu}(2022)}]{Komatsu:2022nvu}%
  \BibitemOpen
  \bibfield  {author} {\bibinfo {author} {\bibfnamefont {E.}~\bibnamefont
  {Komatsu}},\ }\href@noop {} {\  (\bibinfo {year} {2022})},\ \Eprint
  {http://arxiv.org/abs/2202.13919} {arXiv:2202.13919 [astro-ph.CO]}
  \BibitemShut {NoStop}%
\bibitem [{\citenamefont {Chang}\ \emph {et~al.}(2022)\citenamefont {Chang}
  \emph {et~al.}}]{Chang:2022tzj}%
  \BibitemOpen
  \bibfield  {author} {\bibinfo {author} {\bibfnamefont {C.~L.}\ \bibnamefont
  {Chang}} \emph {et~al.},\ }\href@noop {} {\  (\bibinfo {year} {2022})},\
  \Eprint {http://arxiv.org/abs/2203.07638} {arXiv:2203.07638 [astro-ph.CO]}
  \BibitemShut {NoStop}%
\bibitem [{\citenamefont {Sarkar}(1996)}]{Sarkar:1995dd}%
  \BibitemOpen
  \bibfield  {author} {\bibinfo {author} {\bibfnamefont {S.}~\bibnamefont
  {Sarkar}},\ }\href {\doibase 10.1088/0034-4885/59/12/001} {\bibfield
  {journal} {\bibinfo  {journal} {Rept. Prog. Phys.}\ }\textbf {\bibinfo
  {volume} {59}},\ \bibinfo {pages} {1493} (\bibinfo {year} {1996})},\ \Eprint
  {http://arxiv.org/abs/hep-ph/9602260} {arXiv:hep-ph/9602260} \BibitemShut
  {NoStop}%
\bibitem [{\citenamefont {Steigman}(2007)}]{Steigman:2007xt}%
  \BibitemOpen
  \bibfield  {author} {\bibinfo {author} {\bibfnamefont {G.}~\bibnamefont
  {Steigman}},\ }\href {\doibase 10.1146/annurev.nucl.56.080805.140437}
  {\bibfield  {journal} {\bibinfo  {journal} {Ann. Rev. Nucl. Part. Sci.}\
  }\textbf {\bibinfo {volume} {57}},\ \bibinfo {pages} {463} (\bibinfo {year}
  {2007})},\ \Eprint {http://arxiv.org/abs/0712.1100} {arXiv:0712.1100
  [astro-ph]} \BibitemShut {NoStop}%
\bibitem [{\citenamefont {Iocco}\ \emph {et~al.}(2009)\citenamefont {Iocco},
  \citenamefont {Mangano}, \citenamefont {Miele}, \citenamefont {Pisanti},\
  and\ \citenamefont {Serpico}}]{Iocco:2008va}%
  \BibitemOpen
  \bibfield  {author} {\bibinfo {author} {\bibfnamefont {F.}~\bibnamefont
  {Iocco}}, \bibinfo {author} {\bibfnamefont {G.}~\bibnamefont {Mangano}},
  \bibinfo {author} {\bibfnamefont {G.}~\bibnamefont {Miele}}, \bibinfo
  {author} {\bibfnamefont {O.}~\bibnamefont {Pisanti}}, \ and\ \bibinfo
  {author} {\bibfnamefont {P.~D.}\ \bibnamefont {Serpico}},\ }\href {\doibase
  10.1016/j.physrep.2009.02.002} {\bibfield  {journal} {\bibinfo  {journal}
  {Phys. Rept.}\ }\textbf {\bibinfo {volume} {472}},\ \bibinfo {pages} {1}
  (\bibinfo {year} {2009})},\ \Eprint {http://arxiv.org/abs/0809.0631}
  {arXiv:0809.0631 [astro-ph]} \BibitemShut {NoStop}%
\bibitem [{\citenamefont {Pospelov}\ and\ \citenamefont
  {Pradler}(2010)}]{Pospelov:2010hj}%
  \BibitemOpen
  \bibfield  {author} {\bibinfo {author} {\bibfnamefont {M.}~\bibnamefont
  {Pospelov}}\ and\ \bibinfo {author} {\bibfnamefont {J.}~\bibnamefont
  {Pradler}},\ }\href {\doibase 10.1146/annurev.nucl.012809.104521} {\bibfield
  {journal} {\bibinfo  {journal} {Ann. Rev. Nucl. Part. Sci.}\ }\textbf
  {\bibinfo {volume} {60}},\ \bibinfo {pages} {539} (\bibinfo {year} {2010})},\
  \Eprint {http://arxiv.org/abs/1011.1054} {arXiv:1011.1054 [hep-ph]}
  \BibitemShut {NoStop}%
\bibitem [{\citenamefont {Serpico}\ and\ \citenamefont
  {Raffelt}(2004)}]{Serpico:2004nm}%
  \BibitemOpen
  \bibfield  {author} {\bibinfo {author} {\bibfnamefont {P.~D.}\ \bibnamefont
  {Serpico}}\ and\ \bibinfo {author} {\bibfnamefont {G.~G.}\ \bibnamefont
  {Raffelt}},\ }\href {\doibase 10.1103/PhysRevD.70.043526} {\bibfield
  {journal} {\bibinfo  {journal} {Phys. Rev. D}\ }\textbf {\bibinfo {volume}
  {70}},\ \bibinfo {pages} {043526} (\bibinfo {year} {2004})},\ \Eprint
  {http://arxiv.org/abs/astro-ph/0403417} {arXiv:astro-ph/0403417} \BibitemShut
  {NoStop}%
\bibitem [{\citenamefont {Boehm}\ \emph {et~al.}(2013)\citenamefont {Boehm},
  \citenamefont {Dolan},\ and\ \citenamefont {McCabe}}]{Boehm:2013jpa}%
  \BibitemOpen
  \bibfield  {author} {\bibinfo {author} {\bibfnamefont {C.}~\bibnamefont
  {Boehm}}, \bibinfo {author} {\bibfnamefont {M.~J.}\ \bibnamefont {Dolan}}, \
  and\ \bibinfo {author} {\bibfnamefont {C.}~\bibnamefont {McCabe}},\ }\href
  {\doibase 10.1088/1475-7516/2013/08/041} {\bibfield  {journal} {\bibinfo
  {journal} {JCAP}\ }\textbf {\bibinfo {volume} {08}},\ \bibinfo {pages} {041}
  (\bibinfo {year} {2013})},\ \Eprint {http://arxiv.org/abs/1303.6270}
  {arXiv:1303.6270 [hep-ph]} \BibitemShut {NoStop}%
\bibitem [{\citenamefont {Sabti}\ \emph {et~al.}(2020)\citenamefont {Sabti},
  \citenamefont {Alvey}, \citenamefont {Escudero}, \citenamefont {Fairbairn},\
  and\ \citenamefont {Blas}}]{Sabti:2019mhn}%
  \BibitemOpen
  \bibfield  {author} {\bibinfo {author} {\bibfnamefont {N.}~\bibnamefont
  {Sabti}}, \bibinfo {author} {\bibfnamefont {J.}~\bibnamefont {Alvey}},
  \bibinfo {author} {\bibfnamefont {M.}~\bibnamefont {Escudero}}, \bibinfo
  {author} {\bibfnamefont {M.}~\bibnamefont {Fairbairn}}, \ and\ \bibinfo
  {author} {\bibfnamefont {D.}~\bibnamefont {Blas}},\ }\href {\doibase
  10.1088/1475-7516/2020/01/004} {\bibfield  {journal} {\bibinfo  {journal}
  {JCAP}\ }\textbf {\bibinfo {volume} {01}},\ \bibinfo {pages} {004} (\bibinfo
  {year} {2020})},\ \Eprint {http://arxiv.org/abs/1910.01649} {arXiv:1910.01649
  [hep-ph]} \BibitemShut {NoStop}%
\bibitem [{\citenamefont {Giovanetti}\ \emph {et~al.}(2021)\citenamefont
  {Giovanetti}, \citenamefont {Lisanti}, \citenamefont {Liu},\ and\
  \citenamefont {Ruderman}}]{Giovanetti:2021izc}%
  \BibitemOpen
  \bibfield  {author} {\bibinfo {author} {\bibfnamefont {C.}~\bibnamefont
  {Giovanetti}}, \bibinfo {author} {\bibfnamefont {M.}~\bibnamefont {Lisanti}},
  \bibinfo {author} {\bibfnamefont {H.}~\bibnamefont {Liu}}, \ and\ \bibinfo
  {author} {\bibfnamefont {J.~T.}\ \bibnamefont {Ruderman}},\ }\href@noop {} {\
   (\bibinfo {year} {2021})},\ \Eprint {http://arxiv.org/abs/2109.03246}
  {arXiv:2109.03246 [hep-ph]} \BibitemShut {NoStop}%
\bibitem [{\citenamefont {Cooke}\ \emph {et~al.}(2016)\citenamefont {Cooke},
  \citenamefont {Pettini}, \citenamefont {Nollett},\ and\ \citenamefont
  {Jorgenson}}]{Cooke:2016rky}%
  \BibitemOpen
  \bibfield  {author} {\bibinfo {author} {\bibfnamefont {R.~J.}\ \bibnamefont
  {Cooke}}, \bibinfo {author} {\bibfnamefont {M.}~\bibnamefont {Pettini}},
  \bibinfo {author} {\bibfnamefont {K.~M.}\ \bibnamefont {Nollett}}, \ and\
  \bibinfo {author} {\bibfnamefont {R.}~\bibnamefont {Jorgenson}},\ }\href
  {\doibase 10.3847/0004-637X/830/2/148} {\bibfield  {journal} {\bibinfo
  {journal} {Astrophys. J.}\ }\textbf {\bibinfo {volume} {830}},\ \bibinfo
  {pages} {148} (\bibinfo {year} {2016})},\ \Eprint
  {http://arxiv.org/abs/1607.03900} {arXiv:1607.03900 [astro-ph.CO]}
  \BibitemShut {NoStop}%
\bibitem [{\citenamefont {Riemer-S\o{}rensen}\ \emph
  {et~al.}(2017)\citenamefont {Riemer-S\o{}rensen}, \citenamefont {Kotu\v{s}},
  \citenamefont {Webb}, \citenamefont {Ali}, \citenamefont {Dumont},
  \citenamefont {Murphy},\ and\ \citenamefont
  {Carswell}}]{Riemer-Sorensen:2017pey}%
  \BibitemOpen
  \bibfield  {author} {\bibinfo {author} {\bibfnamefont {S.}~\bibnamefont
  {Riemer-S\o{}rensen}}, \bibinfo {author} {\bibfnamefont {S.}~\bibnamefont
  {Kotu\v{s}}}, \bibinfo {author} {\bibfnamefont {J.~K.}\ \bibnamefont {Webb}},
  \bibinfo {author} {\bibfnamefont {K.}~\bibnamefont {Ali}}, \bibinfo {author}
  {\bibfnamefont {V.}~\bibnamefont {Dumont}}, \bibinfo {author} {\bibfnamefont
  {M.~T.}\ \bibnamefont {Murphy}}, \ and\ \bibinfo {author} {\bibfnamefont
  {R.~F.}\ \bibnamefont {Carswell}},\ }\href {\doibase 10.1093/mnras/stx681}
  {\bibfield  {journal} {\bibinfo  {journal} {Mon. Not. Roy. Astron. Soc.}\
  }\textbf {\bibinfo {volume} {468}},\ \bibinfo {pages} {3239} (\bibinfo {year}
  {2017})},\ \Eprint {http://arxiv.org/abs/1703.06656} {arXiv:1703.06656
  [astro-ph.CO]} \BibitemShut {NoStop}%
\bibitem [{\citenamefont {Cooke}\ \emph {et~al.}(2018)\citenamefont {Cooke},
  \citenamefont {Pettini},\ and\ \citenamefont {Steidel}}]{Cooke:2017cwo}%
  \BibitemOpen
  \bibfield  {author} {\bibinfo {author} {\bibfnamefont {R.~J.}\ \bibnamefont
  {Cooke}}, \bibinfo {author} {\bibfnamefont {M.}~\bibnamefont {Pettini}}, \
  and\ \bibinfo {author} {\bibfnamefont {C.~C.}\ \bibnamefont {Steidel}},\
  }\href {\doibase 10.3847/1538-4357/aaab53} {\bibfield  {journal} {\bibinfo
  {journal} {Astrophys. J.}\ }\textbf {\bibinfo {volume} {855}},\ \bibinfo
  {pages} {102} (\bibinfo {year} {2018})},\ \Eprint
  {http://arxiv.org/abs/1710.11129} {arXiv:1710.11129 [astro-ph.CO]}
  \BibitemShut {NoStop}%
\bibitem [{\citenamefont {Zyla}\ \emph {et~al.}(2020)\citenamefont {Zyla} \emph
  {et~al.}}]{Zyla:2020zbs}%
  \BibitemOpen
  \bibfield  {author} {\bibinfo {author} {\bibfnamefont {P.}~\bibnamefont
  {Zyla}} \emph {et~al.} (\bibinfo {collaboration} {Particle Data Group}),\
  }\href {\doibase 10.1093/ptep/ptaa104} {\bibfield  {journal} {\bibinfo
  {journal} {PTEP}\ }\textbf {\bibinfo {volume} {2020}},\ \bibinfo {pages}
  {083C01} (\bibinfo {year} {2020})},\ \bibinfo {note} {and 2021
  update}\BibitemShut {NoStop}%
\bibitem [{\citenamefont {Pitrou}\ \emph
  {et~al.}(2021{\natexlab{a}})\citenamefont {Pitrou}, \citenamefont {Coc},
  \citenamefont {Uzan},\ and\ \citenamefont {Vangioni}}]{Pitrou:2020etk}%
  \BibitemOpen
  \bibfield  {author} {\bibinfo {author} {\bibfnamefont {C.}~\bibnamefont
  {Pitrou}}, \bibinfo {author} {\bibfnamefont {A.}~\bibnamefont {Coc}},
  \bibinfo {author} {\bibfnamefont {J.-P.}\ \bibnamefont {Uzan}}, \ and\
  \bibinfo {author} {\bibfnamefont {E.}~\bibnamefont {Vangioni}},\ }\href
  {\doibase 10.1093/mnras/stab135} {\bibfield  {journal} {\bibinfo  {journal}
  {Mon. Not. Roy. Astron. Soc.}\ }\textbf {\bibinfo {volume} {502}},\ \bibinfo
  {pages} {2474} (\bibinfo {year} {2021}{\natexlab{a}})},\ \Eprint
  {http://arxiv.org/abs/2011.11320} {arXiv:2011.11320 [astro-ph.CO]}
  \BibitemShut {NoStop}%
\bibitem [{\citenamefont {Pisanti}\ \emph {et~al.}(2021)\citenamefont
  {Pisanti}, \citenamefont {Mangano}, \citenamefont {Miele},\ and\
  \citenamefont {Mazzella}}]{Pisanti:2020efz}%
  \BibitemOpen
  \bibfield  {author} {\bibinfo {author} {\bibfnamefont {O.}~\bibnamefont
  {Pisanti}}, \bibinfo {author} {\bibfnamefont {G.}~\bibnamefont {Mangano}},
  \bibinfo {author} {\bibfnamefont {G.}~\bibnamefont {Miele}}, \ and\ \bibinfo
  {author} {\bibfnamefont {P.}~\bibnamefont {Mazzella}},\ }\href {\doibase
  10.1088/1475-7516/2021/04/020} {\bibfield  {journal} {\bibinfo  {journal}
  {JCAP}\ }\textbf {\bibinfo {volume} {04}},\ \bibinfo {pages} {020} (\bibinfo
  {year} {2021})},\ \Eprint {http://arxiv.org/abs/2011.11537} {arXiv:2011.11537
  [astro-ph.CO]} \BibitemShut {NoStop}%
\bibitem [{\citenamefont {Yeh}\ \emph {et~al.}(2021)\citenamefont {Yeh},
  \citenamefont {Olive},\ and\ \citenamefont {Fields}}]{Yeh:2020mgl}%
  \BibitemOpen
  \bibfield  {author} {\bibinfo {author} {\bibfnamefont {T.-H.}\ \bibnamefont
  {Yeh}}, \bibinfo {author} {\bibfnamefont {K.~A.}\ \bibnamefont {Olive}}, \
  and\ \bibinfo {author} {\bibfnamefont {B.~D.}\ \bibnamefont {Fields}},\
  }\href {\doibase 10.1088/1475-7516/2021/03/046} {\bibfield  {journal}
  {\bibinfo  {journal} {JCAP}\ }\textbf {\bibinfo {volume} {03}},\ \bibinfo
  {pages} {046} (\bibinfo {year} {2021})},\ \Eprint
  {http://arxiv.org/abs/2011.13874} {arXiv:2011.13874 [astro-ph.CO]}
  \BibitemShut {NoStop}%
\bibitem [{\citenamefont {Pitrou}\ \emph
  {et~al.}(2021{\natexlab{b}})\citenamefont {Pitrou}, \citenamefont {Coc},
  \citenamefont {Uzan},\ and\ \citenamefont {Vangioni}}]{Pitrou:2021vqr}%
  \BibitemOpen
  \bibfield  {author} {\bibinfo {author} {\bibfnamefont {C.}~\bibnamefont
  {Pitrou}}, \bibinfo {author} {\bibfnamefont {A.}~\bibnamefont {Coc}},
  \bibinfo {author} {\bibfnamefont {J.-P.}\ \bibnamefont {Uzan}}, \ and\
  \bibinfo {author} {\bibfnamefont {E.}~\bibnamefont {Vangioni}},\ }\href
  {\doibase 10.1038/s42254-021-00294-6} {\bibfield  {journal} {\bibinfo
  {journal} {Nature Rev. Phys.}\ }\textbf {\bibinfo {volume} {3}},\ \bibinfo
  {pages} {231} (\bibinfo {year} {2021}{\natexlab{b}})},\ \Eprint
  {http://arxiv.org/abs/2104.11148} {arXiv:2104.11148 [astro-ph.CO]}
  \BibitemShut {NoStop}%
\bibitem [{\citenamefont {Mossa}\ \emph {et~al.}(2020)\citenamefont {Mossa}
  \emph {et~al.}}]{Mossa:2020gjc}%
  \BibitemOpen
  \bibfield  {author} {\bibinfo {author} {\bibfnamefont {V.}~\bibnamefont
  {Mossa}} \emph {et~al.},\ }\href {\doibase 10.1038/s41586-020-2878-4}
  {\bibfield  {journal} {\bibinfo  {journal} {Nature}\ }\textbf {\bibinfo
  {volume} {587}},\ \bibinfo {pages} {210} (\bibinfo {year}
  {2020})}\BibitemShut {NoStop}%
\bibitem [{\citenamefont {Sabti}\ \emph {et~al.}(2021)\citenamefont {Sabti},
  \citenamefont {Alvey}, \citenamefont {Escudero}, \citenamefont {Fairbairn},\
  and\ \citenamefont {Blas}}]{Sabti:2021reh}%
  \BibitemOpen
  \bibfield  {author} {\bibinfo {author} {\bibfnamefont {N.}~\bibnamefont
  {Sabti}}, \bibinfo {author} {\bibfnamefont {J.}~\bibnamefont {Alvey}},
  \bibinfo {author} {\bibfnamefont {M.}~\bibnamefont {Escudero}}, \bibinfo
  {author} {\bibfnamefont {M.}~\bibnamefont {Fairbairn}}, \ and\ \bibinfo
  {author} {\bibfnamefont {D.}~\bibnamefont {Blas}},\ }\href {\doibase
  10.1088/1475-7516/2021/08/A01} {\bibfield  {journal} {\bibinfo  {journal}
  {JCAP}\ }\textbf {\bibinfo {volume} {08}},\ \bibinfo {pages} {A01} (\bibinfo
  {year} {2021})},\ \Eprint {http://arxiv.org/abs/2107.11232} {arXiv:2107.11232
  [hep-ph]} \BibitemShut {NoStop}%
\bibitem [{\citenamefont {Matsumoto}\ \emph {et~al.}(2022)\citenamefont
  {Matsumoto} \emph {et~al.}}]{Matsumoto:2022tlr}%
  \BibitemOpen
  \bibfield  {author} {\bibinfo {author} {\bibfnamefont {A.}~\bibnamefont
  {Matsumoto}} \emph {et~al.},\ }\href {\doibase 10.3847/1538-4357/ac9ea1}
  {\bibfield  {journal} {\bibinfo  {journal} {Astrophys. J.}\ }\textbf
  {\bibinfo {volume} {941}},\ \bibinfo {pages} {167} (\bibinfo {year}
  {2022})},\ \Eprint {http://arxiv.org/abs/2203.09617} {arXiv:2203.09617
  [astro-ph.CO]} \BibitemShut {NoStop}%
\bibitem [{\citenamefont {{Hsyu}}\ \emph {et~al.}(2020)\citenamefont {{Hsyu}},
  \citenamefont {{Cooke}}, \citenamefont {{Prochaska}},\ and\ \citenamefont
  {{Bolte}}}]{2020ApJ...896...77H}%
  \BibitemOpen
  \bibfield  {author} {\bibinfo {author} {\bibfnamefont {T.}~\bibnamefont
  {{Hsyu}}}, \bibinfo {author} {\bibfnamefont {R.~J.}\ \bibnamefont {{Cooke}}},
  \bibinfo {author} {\bibfnamefont {J.~X.}\ \bibnamefont {{Prochaska}}}, \ and\
  \bibinfo {author} {\bibfnamefont {M.}~\bibnamefont {{Bolte}}},\ }\href
  {\doibase 10.3847/1538-4357/ab91af} {\bibfield  {journal} {\bibinfo
  {journal} {\apj}\ }\textbf {\bibinfo {volume} {896}},\ \bibinfo {eid} {77}
  (\bibinfo {year} {2020})},\ \Eprint {http://arxiv.org/abs/2005.12290}
  {arXiv:2005.12290 [astro-ph.GA]} \BibitemShut {NoStop}%
\bibitem [{\citenamefont {Aver}\ \emph {et~al.}(2015)\citenamefont {Aver},
  \citenamefont {Olive},\ and\ \citenamefont {Skillman}}]{Aver:2015iza}%
  \BibitemOpen
  \bibfield  {author} {\bibinfo {author} {\bibfnamefont {E.}~\bibnamefont
  {Aver}}, \bibinfo {author} {\bibfnamefont {K.~A.}\ \bibnamefont {Olive}}, \
  and\ \bibinfo {author} {\bibfnamefont {E.~D.}\ \bibnamefont {Skillman}},\
  }\href {\doibase 10.1088/1475-7516/2015/07/011} {\bibfield  {journal}
  {\bibinfo  {journal} {JCAP}\ }\textbf {\bibinfo {volume} {07}},\ \bibinfo
  {pages} {011} (\bibinfo {year} {2015})},\ \Eprint
  {http://arxiv.org/abs/1503.08146} {arXiv:1503.08146 [astro-ph.CO]}
  \BibitemShut {NoStop}%
\bibitem [{Note1()}]{Note1}%
  \BibitemOpen
  \bibinfo {note} {While the methodology developed in this work stands out as a
  robust recipe for an advanced statistical analysis of Early Universe data,
  the measurement published in~\cite {Matsumoto:2022tlr}, focus of the present
  study, strongly depends on the emission-line data modeling. Further scrutiny
  on the corresponding systematics of the measurement is warranted for the
  future.}\BibitemShut {Stop}%
\bibitem [{\citenamefont {{A.-K.~Burns,~T.~M.P.~Tait,~and~M.~Valli}}()}]{PRyM}%
  \BibitemOpen
  \bibfield  {author} {\bibinfo {author} {\bibnamefont
  {{A.-K.~Burns,~T.~M.P.~Tait,~and~M.~Valli}}},\ }\href@noop {} {\enquote
  {\bibinfo {title} {\texttt{P{R}y{M}ordial}: The first minutes beyond the
  {S}tandard {M}odel in a few seconds},}\ }\bibinfo {howpublished} {\textit{in
  preparation}}\BibitemShut {NoStop}%
\bibitem [{\citenamefont {Kolb}\ and\ \citenamefont
  {Turner}(1990)}]{Kolb:1990vq}%
  \BibitemOpen
  \bibfield  {author} {\bibinfo {author} {\bibfnamefont {E.~W.}\ \bibnamefont
  {Kolb}}\ and\ \bibinfo {author} {\bibfnamefont {M.~S.}\ \bibnamefont
  {Turner}},\ }\href {\doibase 10.1201/9780429492860} {\emph {\bibinfo {title}
  {{The Early Universe}}}},\ Vol.~\bibinfo {volume} {69}\ (\bibinfo {year}
  {1990})\BibitemShut {NoStop}%
\bibitem [{\citenamefont {Rubakov}\ and\ \citenamefont
  {Gorbunov}(2017)}]{Rubakov:2017xzr}%
  \BibitemOpen
  \bibfield  {author} {\bibinfo {author} {\bibfnamefont {V.~A.}\ \bibnamefont
  {Rubakov}}\ and\ \bibinfo {author} {\bibfnamefont {D.~S.}\ \bibnamefont
  {Gorbunov}},\ }\href {\doibase 10.1142/10447} {\emph {\bibinfo {title}
  {{Introduction to the Theory of the Early Universe}: {Hot big bang
  theory}}}}\ (\bibinfo  {publisher} {World Scientific},\ \bibinfo {address}
  {Singapore},\ \bibinfo {year} {2017})\BibitemShut {NoStop}%
\bibitem [{\citenamefont {Simha}\ and\ \citenamefont
  {Steigman}(2008)}]{Simha:2008mt}%
  \BibitemOpen
  \bibfield  {author} {\bibinfo {author} {\bibfnamefont {V.}~\bibnamefont
  {Simha}}\ and\ \bibinfo {author} {\bibfnamefont {G.}~\bibnamefont
  {Steigman}},\ }\href {\doibase 10.1088/1475-7516/2008/08/011} {\bibfield
  {journal} {\bibinfo  {journal} {JCAP}\ }\textbf {\bibinfo {volume} {08}},\
  \bibinfo {pages} {011} (\bibinfo {year} {2008})},\ \Eprint
  {http://arxiv.org/abs/0806.0179} {arXiv:0806.0179 [hep-ph]} \BibitemShut
  {NoStop}%
\bibitem [{\citenamefont {Akita}\ and\ \citenamefont
  {Yamaguchi}(2020)}]{Akita:2020szl}%
  \BibitemOpen
  \bibfield  {author} {\bibinfo {author} {\bibfnamefont {K.}~\bibnamefont
  {Akita}}\ and\ \bibinfo {author} {\bibfnamefont {M.}~\bibnamefont
  {Yamaguchi}},\ }\href {\doibase 10.1088/1475-7516/2020/08/012} {\bibfield
  {journal} {\bibinfo  {journal} {JCAP}\ }\textbf {\bibinfo {volume} {08}},\
  \bibinfo {pages} {012} (\bibinfo {year} {2020})},\ \Eprint
  {http://arxiv.org/abs/2005.07047} {arXiv:2005.07047 [hep-ph]} \BibitemShut
  {NoStop}%
\bibitem [{\citenamefont {Froustey}\ \emph {et~al.}(2020)\citenamefont
  {Froustey}, \citenamefont {Pitrou},\ and\ \citenamefont
  {Volpe}}]{Froustey:2020mcq}%
  \BibitemOpen
  \bibfield  {author} {\bibinfo {author} {\bibfnamefont {J.}~\bibnamefont
  {Froustey}}, \bibinfo {author} {\bibfnamefont {C.}~\bibnamefont {Pitrou}}, \
  and\ \bibinfo {author} {\bibfnamefont {M.~C.}\ \bibnamefont {Volpe}},\ }\href
  {\doibase 10.1088/1475-7516/2020/12/015} {\bibfield  {journal} {\bibinfo
  {journal} {JCAP}\ }\textbf {\bibinfo {volume} {12}},\ \bibinfo {pages} {015}
  (\bibinfo {year} {2020})},\ \Eprint {http://arxiv.org/abs/2008.01074}
  {arXiv:2008.01074 [hep-ph]} \BibitemShut {NoStop}%
\bibitem [{\citenamefont {Bennett}\ \emph {et~al.}(2021)\citenamefont
  {Bennett}, \citenamefont {Buldgen}, \citenamefont {De~Salas}, \citenamefont
  {Drewes}, \citenamefont {Gariazzo}, \citenamefont {Pastor},\ and\
  \citenamefont {Wong}}]{Bennett:2020zkv}%
  \BibitemOpen
  \bibfield  {author} {\bibinfo {author} {\bibfnamefont {J.~J.}\ \bibnamefont
  {Bennett}}, \bibinfo {author} {\bibfnamefont {G.}~\bibnamefont {Buldgen}},
  \bibinfo {author} {\bibfnamefont {P.~F.}\ \bibnamefont {De~Salas}}, \bibinfo
  {author} {\bibfnamefont {M.}~\bibnamefont {Drewes}}, \bibinfo {author}
  {\bibfnamefont {S.}~\bibnamefont {Gariazzo}}, \bibinfo {author}
  {\bibfnamefont {S.}~\bibnamefont {Pastor}}, \ and\ \bibinfo {author}
  {\bibfnamefont {Y.~Y.~Y.}\ \bibnamefont {Wong}},\ }\href {\doibase
  10.1088/1475-7516/2021/04/073} {\bibfield  {journal} {\bibinfo  {journal}
  {JCAP}\ }\textbf {\bibinfo {volume} {04}},\ \bibinfo {pages} {073} (\bibinfo
  {year} {2021})},\ \Eprint {http://arxiv.org/abs/2012.02726} {arXiv:2012.02726
  [hep-ph]} \BibitemShut {NoStop}%
\bibitem [{\citenamefont {Grohs}\ \emph {et~al.}(2016)\citenamefont {Grohs},
  \citenamefont {Fuller}, \citenamefont {Kishimoto}, \citenamefont {Paris},\
  and\ \citenamefont {Vlasenko}}]{Grohs:2015tfy}%
  \BibitemOpen
  \bibfield  {author} {\bibinfo {author} {\bibfnamefont {E.}~\bibnamefont
  {Grohs}}, \bibinfo {author} {\bibfnamefont {G.~M.}\ \bibnamefont {Fuller}},
  \bibinfo {author} {\bibfnamefont {C.~T.}\ \bibnamefont {Kishimoto}}, \bibinfo
  {author} {\bibfnamefont {M.~W.}\ \bibnamefont {Paris}}, \ and\ \bibinfo
  {author} {\bibfnamefont {A.}~\bibnamefont {Vlasenko}},\ }\href {\doibase
  10.1103/PhysRevD.93.083522} {\bibfield  {journal} {\bibinfo  {journal} {Phys.
  Rev. D}\ }\textbf {\bibinfo {volume} {93}},\ \bibinfo {pages} {083522}
  (\bibinfo {year} {2016})},\ \Eprint {http://arxiv.org/abs/1512.02205}
  {arXiv:1512.02205 [astro-ph.CO]} \BibitemShut {NoStop}%
\bibitem [{\citenamefont {Escudero~Abenza}(2020)}]{EscuderoAbenza:2020cmq}%
  \BibitemOpen
  \bibfield  {author} {\bibinfo {author} {\bibfnamefont {M.}~\bibnamefont
  {Escudero~Abenza}},\ }\href {\doibase 10.1088/1475-7516/2020/05/048}
  {\bibfield  {journal} {\bibinfo  {journal} {JCAP}\ }\textbf {\bibinfo
  {volume} {05}},\ \bibinfo {pages} {048} (\bibinfo {year} {2020})},\ \Eprint
  {http://arxiv.org/abs/2001.04466} {arXiv:2001.04466 [hep-ph]} \BibitemShut
  {NoStop}%
\bibitem [{\citenamefont {Kinney}\ and\ \citenamefont
  {Riotto}(1999)}]{Kinney:1999pd}%
  \BibitemOpen
  \bibfield  {author} {\bibinfo {author} {\bibfnamefont {W.~H.}\ \bibnamefont
  {Kinney}}\ and\ \bibinfo {author} {\bibfnamefont {A.}~\bibnamefont
  {Riotto}},\ }\href {\doibase 10.1103/PhysRevLett.83.3366} {\bibfield
  {journal} {\bibinfo  {journal} {Phys. Rev. Lett.}\ }\textbf {\bibinfo
  {volume} {83}},\ \bibinfo {pages} {3366} (\bibinfo {year} {1999})},\ \Eprint
  {http://arxiv.org/abs/hep-ph/9903459} {arXiv:hep-ph/9903459} \BibitemShut
  {NoStop}%
\bibitem [{\citenamefont {Lesgourgues}\ and\ \citenamefont
  {Pastor}(1999)}]{Lesgourgues:1999wu}%
  \BibitemOpen
  \bibfield  {author} {\bibinfo {author} {\bibfnamefont {J.}~\bibnamefont
  {Lesgourgues}}\ and\ \bibinfo {author} {\bibfnamefont {S.}~\bibnamefont
  {Pastor}},\ }\href {\doibase 10.1103/PhysRevD.60.103521} {\bibfield
  {journal} {\bibinfo  {journal} {Phys. Rev. D}\ }\textbf {\bibinfo {volume}
  {60}},\ \bibinfo {pages} {103521} (\bibinfo {year} {1999})},\ \Eprint
  {http://arxiv.org/abs/hep-ph/9904411} {arXiv:hep-ph/9904411} \BibitemShut
  {NoStop}%
\bibitem [{\citenamefont {Oldengott}\ and\ \citenamefont
  {Schwarz}(2017)}]{Oldengott:2017tzj}%
  \BibitemOpen
  \bibfield  {author} {\bibinfo {author} {\bibfnamefont {I.~M.}\ \bibnamefont
  {Oldengott}}\ and\ \bibinfo {author} {\bibfnamefont {D.~J.}\ \bibnamefont
  {Schwarz}},\ }\href {\doibase 10.1209/0295-5075/119/29001} {\bibfield
  {journal} {\bibinfo  {journal} {EPL}\ }\textbf {\bibinfo {volume} {119}},\
  \bibinfo {pages} {29001} (\bibinfo {year} {2017})},\ \Eprint
  {http://arxiv.org/abs/1706.01705} {arXiv:1706.01705 [astro-ph.CO]}
  \BibitemShut {NoStop}%
\bibitem [{\citenamefont {Bonilla}\ \emph {et~al.}(2019)\citenamefont
  {Bonilla}, \citenamefont {Nunes},\ and\ \citenamefont
  {Abreu}}]{Bonilla:2018nau}%
  \BibitemOpen
  \bibfield  {author} {\bibinfo {author} {\bibfnamefont {A.}~\bibnamefont
  {Bonilla}}, \bibinfo {author} {\bibfnamefont {R.~C.}\ \bibnamefont {Nunes}},
  \ and\ \bibinfo {author} {\bibfnamefont {E.~M.~C.}\ \bibnamefont {Abreu}},\
  }\href {\doibase 10.1093/mnras/stz524} {\bibfield  {journal} {\bibinfo
  {journal} {Mon. Not. Roy. Astron. Soc.}\ }\textbf {\bibinfo {volume} {485}},\
  \bibinfo {pages} {2486} (\bibinfo {year} {2019})},\ \Eprint
  {http://arxiv.org/abs/1810.06356} {arXiv:1810.06356 [astro-ph.CO]}
  \BibitemShut {NoStop}%
\bibitem [{\citenamefont {Kumar}\ \emph {et~al.}(2022)\citenamefont {Kumar},
  \citenamefont {Nunes},\ and\ \citenamefont {Yadav}}]{Kumar:2022vee}%
  \BibitemOpen
  \bibfield  {author} {\bibinfo {author} {\bibfnamefont {S.}~\bibnamefont
  {Kumar}}, \bibinfo {author} {\bibfnamefont {R.~C.}\ \bibnamefont {Nunes}}, \
  and\ \bibinfo {author} {\bibfnamefont {P.}~\bibnamefont {Yadav}},\
  }\href@noop {} {\  (\bibinfo {year} {2022})},\ \Eprint
  {http://arxiv.org/abs/2205.04292} {arXiv:2205.04292 [astro-ph.CO]}
  \BibitemShut {NoStop}%
\bibitem [{\citenamefont {Abazajian}\ \emph {et~al.}(2002)\citenamefont
  {Abazajian}, \citenamefont {Beacom},\ and\ \citenamefont
  {Bell}}]{Abazajian:2002qx}%
  \BibitemOpen
  \bibfield  {author} {\bibinfo {author} {\bibfnamefont {K.~N.}\ \bibnamefont
  {Abazajian}}, \bibinfo {author} {\bibfnamefont {J.~F.}\ \bibnamefont
  {Beacom}}, \ and\ \bibinfo {author} {\bibfnamefont {N.~F.}\ \bibnamefont
  {Bell}},\ }\href {\doibase 10.1103/PhysRevD.66.013008} {\bibfield  {journal}
  {\bibinfo  {journal} {Phys. Rev. D}\ }\textbf {\bibinfo {volume} {66}},\
  \bibinfo {pages} {013008} (\bibinfo {year} {2002})},\ \Eprint
  {http://arxiv.org/abs/astro-ph/0203442} {arXiv:astro-ph/0203442} \BibitemShut
  {NoStop}%
\bibitem [{\citenamefont {Castorina}\ \emph {et~al.}(2012)\citenamefont
  {Castorina}, \citenamefont {Franca}, \citenamefont {Lattanzi}, \citenamefont
  {Lesgourgues}, \citenamefont {Mangano}, \citenamefont {Melchiorri},\ and\
  \citenamefont {Pastor}}]{Castorina:2012md}%
  \BibitemOpen
  \bibfield  {author} {\bibinfo {author} {\bibfnamefont {E.}~\bibnamefont
  {Castorina}}, \bibinfo {author} {\bibfnamefont {U.}~\bibnamefont {Franca}},
  \bibinfo {author} {\bibfnamefont {M.}~\bibnamefont {Lattanzi}}, \bibinfo
  {author} {\bibfnamefont {J.}~\bibnamefont {Lesgourgues}}, \bibinfo {author}
  {\bibfnamefont {G.}~\bibnamefont {Mangano}}, \bibinfo {author} {\bibfnamefont
  {A.}~\bibnamefont {Melchiorri}}, \ and\ \bibinfo {author} {\bibfnamefont
  {S.}~\bibnamefont {Pastor}},\ }\href {\doibase 10.1103/PhysRevD.86.023517}
  {\bibfield  {journal} {\bibinfo  {journal} {Phys. Rev. D}\ }\textbf {\bibinfo
  {volume} {86}},\ \bibinfo {pages} {023517} (\bibinfo {year} {2012})},\
  \Eprint {http://arxiv.org/abs/1204.2510} {arXiv:1204.2510 [astro-ph.CO]}
  \BibitemShut {NoStop}%
\bibitem [{\citenamefont {Aghanim}\ \emph
  {et~al.}(2019{\natexlab{a}})\citenamefont {Aghanim} \emph
  {et~al.}}]{Planck18res}%
  \BibitemOpen
  \bibfield  {author} {\bibinfo {author} {\bibfnamefont {N.}~\bibnamefont
  {Aghanim}} \emph {et~al.} (\bibinfo {collaboration} {Planck}),\ }\href@noop
  {} {}\bibinfo {howpublished}
  {\url{https://wiki.cosmos.esa.int/planck-legacy-archive/images/4/43/Baseline_params_table_2018_68pc_v2.pdf}}
  (\bibinfo {year} {2019}{\natexlab{a}})\BibitemShut {NoStop}%
\bibitem [{\citenamefont {Barenboim}\ \emph {et~al.}(2017)\citenamefont
  {Barenboim}, \citenamefont {Kinney},\ and\ \citenamefont
  {Park}}]{Barenboim:2016lxv}%
  \BibitemOpen
  \bibfield  {author} {\bibinfo {author} {\bibfnamefont {G.}~\bibnamefont
  {Barenboim}}, \bibinfo {author} {\bibfnamefont {W.~H.}\ \bibnamefont
  {Kinney}}, \ and\ \bibinfo {author} {\bibfnamefont {W.-I.}\ \bibnamefont
  {Park}},\ }\href {\doibase 10.1140/epjc/s10052-017-5147-4} {\bibfield
  {journal} {\bibinfo  {journal} {Eur. Phys. J. C}\ }\textbf {\bibinfo {volume}
  {77}},\ \bibinfo {pages} {590} (\bibinfo {year} {2017})},\ \Eprint
  {http://arxiv.org/abs/1609.03200} {arXiv:1609.03200 [astro-ph.CO]}
  \BibitemShut {NoStop}%
\bibitem [{\citenamefont {Dolgov}\ \emph {et~al.}(2002)\citenamefont {Dolgov},
  \citenamefont {Hansen}, \citenamefont {Pastor}, \citenamefont {Petcov},
  \citenamefont {Raffelt},\ and\ \citenamefont {Semikoz}}]{Dolgov:2002ab}%
  \BibitemOpen
  \bibfield  {author} {\bibinfo {author} {\bibfnamefont {A.~D.}\ \bibnamefont
  {Dolgov}}, \bibinfo {author} {\bibfnamefont {S.~H.}\ \bibnamefont {Hansen}},
  \bibinfo {author} {\bibfnamefont {S.}~\bibnamefont {Pastor}}, \bibinfo
  {author} {\bibfnamefont {S.~T.}\ \bibnamefont {Petcov}}, \bibinfo {author}
  {\bibfnamefont {G.~G.}\ \bibnamefont {Raffelt}}, \ and\ \bibinfo {author}
  {\bibfnamefont {D.~V.}\ \bibnamefont {Semikoz}},\ }\href {\doibase
  10.1016/S0550-3213(02)00274-2} {\bibfield  {journal} {\bibinfo  {journal}
  {Nucl. Phys. B}\ }\textbf {\bibinfo {volume} {632}},\ \bibinfo {pages} {363}
  (\bibinfo {year} {2002})},\ \Eprint {http://arxiv.org/abs/hep-ph/0201287}
  {arXiv:hep-ph/0201287} \BibitemShut {NoStop}%
\bibitem [{\citenamefont {Pastor}\ \emph {et~al.}(2009)\citenamefont {Pastor},
  \citenamefont {Pinto},\ and\ \citenamefont {Raffelt}}]{Pastor:2008ti}%
  \BibitemOpen
  \bibfield  {author} {\bibinfo {author} {\bibfnamefont {S.}~\bibnamefont
  {Pastor}}, \bibinfo {author} {\bibfnamefont {T.}~\bibnamefont {Pinto}}, \
  and\ \bibinfo {author} {\bibfnamefont {G.~G.}\ \bibnamefont {Raffelt}},\
  }\href {\doibase 10.1103/PhysRevLett.102.241302} {\bibfield  {journal}
  {\bibinfo  {journal} {Phys. Rev. Lett.}\ }\textbf {\bibinfo {volume} {102}},\
  \bibinfo {pages} {241302} (\bibinfo {year} {2009})},\ \Eprint
  {http://arxiv.org/abs/0808.3137} {arXiv:0808.3137 [astro-ph]} \BibitemShut
  {NoStop}%
\bibitem [{\citenamefont {Fields}\ \emph {et~al.}(2020)\citenamefont {Fields},
  \citenamefont {Olive}, \citenamefont {Yeh},\ and\ \citenamefont
  {Young}}]{Fields:2019pfx}%
  \BibitemOpen
  \bibfield  {author} {\bibinfo {author} {\bibfnamefont {B.~D.}\ \bibnamefont
  {Fields}}, \bibinfo {author} {\bibfnamefont {K.~A.}\ \bibnamefont {Olive}},
  \bibinfo {author} {\bibfnamefont {T.-H.}\ \bibnamefont {Yeh}}, \ and\
  \bibinfo {author} {\bibfnamefont {C.}~\bibnamefont {Young}},\ }\href
  {\doibase 10.1088/1475-7516/2020/03/010} {\bibfield  {journal} {\bibinfo
  {journal} {JCAP}\ }\textbf {\bibinfo {volume} {03}},\ \bibinfo {pages} {010}
  (\bibinfo {year} {2020})},\ \bibinfo {note} {[Erratum: JCAP 11, E02
  (2020)]},\ \Eprint {http://arxiv.org/abs/1912.01132} {arXiv:1912.01132
  [astro-ph.CO]} \BibitemShut {NoStop}%
\bibitem [{\citenamefont {Kohri}\ \emph {et~al.}(1997)\citenamefont {Kohri},
  \citenamefont {Kawasaki},\ and\ \citenamefont {Sato}}]{Kohri1997}%
  \BibitemOpen
  \bibfield  {author} {\bibinfo {author} {\bibfnamefont {K.}~\bibnamefont
  {Kohri}}, \bibinfo {author} {\bibfnamefont {M.}~\bibnamefont {Kawasaki}}, \
  and\ \bibinfo {author} {\bibfnamefont {K.}~\bibnamefont {Sato}},\ }\href
  {\doibase 10.1086/512793} {\bibfield  {journal} {\bibinfo  {journal}
  {Astrophys. J.}\ }\textbf {\bibinfo {volume} {490}},\ \bibinfo {pages} {72}
  (\bibinfo {year} {1997})},\ \Eprint {http://arxiv.org/abs/astro-ph/9612237}
  {arXiv:astro-ph/9612237} \BibitemShut {NoStop}%
\bibitem [{\citenamefont {Pitrou}\ \emph {et~al.}(2018)\citenamefont {Pitrou},
  \citenamefont {Coc}, \citenamefont {Uzan},\ and\ \citenamefont
  {Vangioni}}]{Pitrou:2018cgg}%
  \BibitemOpen
  \bibfield  {author} {\bibinfo {author} {\bibfnamefont {C.}~\bibnamefont
  {Pitrou}}, \bibinfo {author} {\bibfnamefont {A.}~\bibnamefont {Coc}},
  \bibinfo {author} {\bibfnamefont {J.-P.}\ \bibnamefont {Uzan}}, \ and\
  \bibinfo {author} {\bibfnamefont {E.}~\bibnamefont {Vangioni}},\ }\href
  {\doibase 10.1016/j.physrep.2018.04.005} {\bibfield  {journal} {\bibinfo
  {journal} {Phys. Rept.}\ }\textbf {\bibinfo {volume} {754}},\ \bibinfo
  {pages} {1} (\bibinfo {year} {2018})},\ \Eprint
  {http://arxiv.org/abs/1801.08023} {arXiv:1801.08023 [astro-ph.CO]}
  \BibitemShut {NoStop}%
\bibitem [{\citenamefont {Froustey}\ and\ \citenamefont
  {Pitrou}(2022)}]{Froustey:2021azz}%
  \BibitemOpen
  \bibfield  {author} {\bibinfo {author} {\bibfnamefont {J.}~\bibnamefont
  {Froustey}}\ and\ \bibinfo {author} {\bibfnamefont {C.}~\bibnamefont
  {Pitrou}},\ }\href {\doibase 10.1088/1475-7516/2022/03/065} {\bibfield
  {journal} {\bibinfo  {journal} {JCAP}\ }\textbf {\bibinfo {volume} {03}},\
  \bibinfo {pages} {065} (\bibinfo {year} {2022})},\ \Eprint
  {http://arxiv.org/abs/2110.11889} {arXiv:2110.11889 [hep-ph]} \BibitemShut
  {NoStop}%
\bibitem [{\citenamefont {Escudero}(2019)}]{Escudero:2018mvt}%
  \BibitemOpen
  \bibfield  {author} {\bibinfo {author} {\bibfnamefont {M.}~\bibnamefont
  {Escudero}},\ }\href {\doibase 10.1088/1475-7516/2019/02/007} {\bibfield
  {journal} {\bibinfo  {journal} {JCAP}\ }\textbf {\bibinfo {volume} {02}},\
  \bibinfo {pages} {007} (\bibinfo {year} {2019})},\ \Eprint
  {http://arxiv.org/abs/1812.05605} {arXiv:1812.05605 [hep-ph]} \BibitemShut
  {NoStop}%
\bibitem [{\citenamefont {Chu}\ \emph {et~al.}(2022)\citenamefont {Chu},
  \citenamefont {Kuo},\ and\ \citenamefont {Pradler}}]{Chu:2022xuh}%
  \BibitemOpen
  \bibfield  {author} {\bibinfo {author} {\bibfnamefont {X.}~\bibnamefont
  {Chu}}, \bibinfo {author} {\bibfnamefont {J.-L.}\ \bibnamefont {Kuo}}, \ and\
  \bibinfo {author} {\bibfnamefont {J.}~\bibnamefont {Pradler}},\ }\href@noop
  {} {\  (\bibinfo {year} {2022})},\ \Eprint {http://arxiv.org/abs/2205.05714}
  {arXiv:2205.05714 [hep-ph]} \BibitemShut {NoStop}%
\bibitem [{\citenamefont {Bennett}\ \emph {et~al.}(2020)\citenamefont
  {Bennett}, \citenamefont {Buldgen}, \citenamefont {Drewes},\ and\
  \citenamefont {Wong}}]{Bennett:2019ewm}%
  \BibitemOpen
  \bibfield  {author} {\bibinfo {author} {\bibfnamefont {J.~J.}\ \bibnamefont
  {Bennett}}, \bibinfo {author} {\bibfnamefont {G.}~\bibnamefont {Buldgen}},
  \bibinfo {author} {\bibfnamefont {M.}~\bibnamefont {Drewes}}, \ and\ \bibinfo
  {author} {\bibfnamefont {Y.~Y.~Y.}\ \bibnamefont {Wong}},\ }\href {\doibase
  10.1088/1475-7516/2020/03/003} {\bibfield  {journal} {\bibinfo  {journal}
  {JCAP}\ }\textbf {\bibinfo {volume} {03}},\ \bibinfo {pages} {003} (\bibinfo
  {year} {2020})},\ \bibinfo {note} {[Addendum: JCAP 03, A01 (2021)]},\ \Eprint
  {http://arxiv.org/abs/1911.04504} {arXiv:1911.04504 [hep-ph]} \BibitemShut
  {NoStop}%
\bibitem [{\citenamefont {Ade}\ \emph {et~al.}(2019)\citenamefont {Ade} \emph
  {et~al.}}]{SimonsObservatory:2018koc}%
  \BibitemOpen
  \bibfield  {author} {\bibinfo {author} {\bibfnamefont {P.}~\bibnamefont
  {Ade}} \emph {et~al.} (\bibinfo {collaboration} {Simons Observatory}),\
  }\href {\doibase 10.1088/1475-7516/2019/02/056} {\bibfield  {journal}
  {\bibinfo  {journal} {JCAP}\ }\textbf {\bibinfo {volume} {02}},\ \bibinfo
  {pages} {056} (\bibinfo {year} {2019})},\ \Eprint
  {http://arxiv.org/abs/1808.07445} {arXiv:1808.07445 [astro-ph.CO]}
  \BibitemShut {NoStop}%
\bibitem [{\citenamefont {Abazajian}\ \emph {et~al.}(2022)\citenamefont
  {Abazajian} \emph {et~al.}}]{CMB-S4:2022ght}%
  \BibitemOpen
  \bibfield  {author} {\bibinfo {author} {\bibfnamefont {K.}~\bibnamefont
  {Abazajian}} \emph {et~al.} (\bibinfo {collaboration} {CMB-S4}),\ }in\
  \href@noop {} {\emph {\bibinfo {booktitle} {{2022 Snowmass Summer Study}}}}\
  (\bibinfo {year} {2022})\ \Eprint {http://arxiv.org/abs/2203.08024}
  {arXiv:2203.08024 [astro-ph.CO]} \BibitemShut {NoStop}%
\bibitem [{\citenamefont {Aiola}\ \emph {et~al.}(2022)\citenamefont {Aiola}
  \emph {et~al.}}]{CMB-HD:2022bsz}%
  \BibitemOpen
  \bibfield  {author} {\bibinfo {author} {\bibfnamefont {S.}~\bibnamefont
  {Aiola}} \emph {et~al.} (\bibinfo {collaboration} {CMB-HD}),\ }\href@noop {}
  {\  (\bibinfo {year} {2022})},\ \Eprint {http://arxiv.org/abs/2203.05728}
  {arXiv:2203.05728 [astro-ph.CO]} \BibitemShut {NoStop}%
\bibitem [{\citenamefont {Sirlin}(1967)}]{Sirlin:1967zza}%
  \BibitemOpen
  \bibfield  {author} {\bibinfo {author} {\bibfnamefont {A.}~\bibnamefont
  {Sirlin}},\ }\href {\doibase 10.1103/PhysRev.164.1767} {\bibfield  {journal}
  {\bibinfo  {journal} {Phys. Rev.}\ }\textbf {\bibinfo {volume} {164}},\
  \bibinfo {pages} {1767} (\bibinfo {year} {1967})}\BibitemShut {NoStop}%
\bibitem [{\citenamefont {Brown}\ and\ \citenamefont
  {Sawyer}(2001)}]{Brown:2000cp}%
  \BibitemOpen
  \bibfield  {author} {\bibinfo {author} {\bibfnamefont {L.~S.}\ \bibnamefont
  {Brown}}\ and\ \bibinfo {author} {\bibfnamefont {R.~F.}\ \bibnamefont
  {Sawyer}},\ }\href {\doibase 10.1103/PhysRevD.63.083503} {\bibfield
  {journal} {\bibinfo  {journal} {Phys. Rev. D}\ }\textbf {\bibinfo {volume}
  {63}},\ \bibinfo {pages} {083503} (\bibinfo {year} {2001})},\ \Eprint
  {http://arxiv.org/abs/astro-ph/0006370} {arXiv:astro-ph/0006370} \BibitemShut
  {NoStop}%
\bibitem [{\citenamefont {Czarnecki}\ \emph {et~al.}(2004)\citenamefont
  {Czarnecki}, \citenamefont {Marciano},\ and\ \citenamefont
  {Sirlin}}]{Czarnecki:2004cw}%
  \BibitemOpen
  \bibfield  {author} {\bibinfo {author} {\bibfnamefont {A.}~\bibnamefont
  {Czarnecki}}, \bibinfo {author} {\bibfnamefont {W.~J.}\ \bibnamefont
  {Marciano}}, \ and\ \bibinfo {author} {\bibfnamefont {A.}~\bibnamefont
  {Sirlin}},\ }\href {\doibase 10.1103/PhysRevD.70.093006} {\bibfield
  {journal} {\bibinfo  {journal} {Phys. Rev. D}\ }\textbf {\bibinfo {volume}
  {70}},\ \bibinfo {pages} {093006} (\bibinfo {year} {2004})},\ \Eprint
  {http://arxiv.org/abs/hep-ph/0406324} {arXiv:hep-ph/0406324} \BibitemShut
  {NoStop}%
\bibitem [{\citenamefont {Ivanov}\ \emph {et~al.}(2017)\citenamefont {Ivanov},
  \citenamefont {H\"ollwieser}, \citenamefont {Troitskaya}, \citenamefont
  {Wellenzohn},\ and\ \citenamefont {Berdnikov}}]{Ivanov:2017fra}%
  \BibitemOpen
  \bibfield  {author} {\bibinfo {author} {\bibfnamefont {A.~N.}\ \bibnamefont
  {Ivanov}}, \bibinfo {author} {\bibfnamefont {R.}~\bibnamefont
  {H\"ollwieser}}, \bibinfo {author} {\bibfnamefont {N.~I.}\ \bibnamefont
  {Troitskaya}}, \bibinfo {author} {\bibfnamefont {M.}~\bibnamefont
  {Wellenzohn}}, \ and\ \bibinfo {author} {\bibfnamefont {Y.~A.}\ \bibnamefont
  {Berdnikov}},\ }\href {\doibase 10.1103/PhysRevD.95.033007} {\bibfield
  {journal} {\bibinfo  {journal} {Phys. Rev. D}\ }\textbf {\bibinfo {volume}
  {95}},\ \bibinfo {pages} {033007} (\bibinfo {year} {2017})},\ \Eprint
  {http://arxiv.org/abs/1701.04613} {arXiv:1701.04613 [hep-ph]} \BibitemShut
  {NoStop}%
\bibitem [{\citenamefont {Lopez}\ \emph {et~al.}(1997)\citenamefont {Lopez},
  \citenamefont {Turner},\ and\ \citenamefont {Gyuk}}]{Lopez:1997ki}%
  \BibitemOpen
  \bibfield  {author} {\bibinfo {author} {\bibfnamefont {R.~E.}\ \bibnamefont
  {Lopez}}, \bibinfo {author} {\bibfnamefont {M.~S.}\ \bibnamefont {Turner}}, \
  and\ \bibinfo {author} {\bibfnamefont {G.}~\bibnamefont {Gyuk}},\ }\href
  {\doibase 10.1103/PhysRevD.56.3191} {\bibfield  {journal} {\bibinfo
  {journal} {Phys. Rev. D}\ }\textbf {\bibinfo {volume} {56}},\ \bibinfo
  {pages} {3191} (\bibinfo {year} {1997})},\ \Eprint
  {http://arxiv.org/abs/astro-ph/9703065} {arXiv:astro-ph/9703065} \BibitemShut
  {NoStop}%
\bibitem [{\citenamefont {Iliadis}\ and\ \citenamefont
  {Coc}(2020)}]{Iliadis:2020jtc}%
  \BibitemOpen
  \bibfield  {author} {\bibinfo {author} {\bibfnamefont {C.}~\bibnamefont
  {Iliadis}}\ and\ \bibinfo {author} {\bibfnamefont {A.}~\bibnamefont {Coc}},\
  }\href {\doibase 10.3847/1538-4357/abb1a3} {\bibfield  {journal} {\bibinfo
  {journal} {Astrophys. J.}\ }\textbf {\bibinfo {volume} {901}},\ \bibinfo
  {pages} {127} (\bibinfo {year} {2020})},\ \Eprint
  {http://arxiv.org/abs/2008.12200} {arXiv:2008.12200 [astro-ph.CO]}
  \BibitemShut {NoStop}%
\bibitem [{\citenamefont {Serpico}\ \emph {et~al.}(2004)\citenamefont
  {Serpico}, \citenamefont {Esposito}, \citenamefont {Iocco}, \citenamefont
  {Mangano}, \citenamefont {Miele},\ and\ \citenamefont
  {Pisanti}}]{Serpico:2004gx}%
  \BibitemOpen
  \bibfield  {author} {\bibinfo {author} {\bibfnamefont {P.~D.}\ \bibnamefont
  {Serpico}}, \bibinfo {author} {\bibfnamefont {S.}~\bibnamefont {Esposito}},
  \bibinfo {author} {\bibfnamefont {F.}~\bibnamefont {Iocco}}, \bibinfo
  {author} {\bibfnamefont {G.}~\bibnamefont {Mangano}}, \bibinfo {author}
  {\bibfnamefont {G.}~\bibnamefont {Miele}}, \ and\ \bibinfo {author}
  {\bibfnamefont {O.}~\bibnamefont {Pisanti}},\ }\href {\doibase
  10.1088/1475-7516/2004/12/010} {\bibfield  {journal} {\bibinfo  {journal}
  {JCAP}\ }\textbf {\bibinfo {volume} {12}},\ \bibinfo {pages} {010} (\bibinfo
  {year} {2004})},\ \Eprint {http://arxiv.org/abs/astro-ph/0408076}
  {arXiv:astro-ph/0408076} \BibitemShut {NoStop}%
\bibitem [{\citenamefont {Ando}\ \emph {et~al.}(2006)\citenamefont {Ando},
  \citenamefont {Cyburt}, \citenamefont {Hong},\ and\ \citenamefont
  {Hyun}}]{Ando:2005cz}%
  \BibitemOpen
  \bibfield  {author} {\bibinfo {author} {\bibfnamefont {S.}~\bibnamefont
  {Ando}}, \bibinfo {author} {\bibfnamefont {R.~H.}\ \bibnamefont {Cyburt}},
  \bibinfo {author} {\bibfnamefont {S.~W.}\ \bibnamefont {Hong}}, \ and\
  \bibinfo {author} {\bibfnamefont {C.~H.}\ \bibnamefont {Hyun}},\ }\href
  {\doibase 10.1103/PhysRevC.74.025809} {\bibfield  {journal} {\bibinfo
  {journal} {Phys. Rev. C}\ }\textbf {\bibinfo {volume} {74}},\ \bibinfo
  {pages} {025809} (\bibinfo {year} {2006})},\ \Eprint
  {http://arxiv.org/abs/nucl-th/0511074} {arXiv:nucl-th/0511074} \BibitemShut
  {NoStop}%
\bibitem [{\citenamefont {Descouvemont}\ \emph {et~al.}(2004)\citenamefont
  {Descouvemont}, \citenamefont {Adahchour}, \citenamefont {Angulo},
  \citenamefont {Coc},\ and\ \citenamefont
  {Vangioni-Flam}}]{Descouvemont:2004cw}%
  \BibitemOpen
  \bibfield  {author} {\bibinfo {author} {\bibfnamefont {P.}~\bibnamefont
  {Descouvemont}}, \bibinfo {author} {\bibfnamefont {A.}~\bibnamefont
  {Adahchour}}, \bibinfo {author} {\bibfnamefont {C.}~\bibnamefont {Angulo}},
  \bibinfo {author} {\bibfnamefont {A.}~\bibnamefont {Coc}}, \ and\ \bibinfo
  {author} {\bibfnamefont {E.}~\bibnamefont {Vangioni-Flam}},\ }\href {\doibase
  10.1016/j.adt.2004.08.001} {\bibfield  {journal} {\bibinfo  {journal} {Atom.
  Data Nucl. Data Tabl.}\ }\textbf {\bibinfo {volume} {88}},\ \bibinfo {pages}
  {203} (\bibinfo {year} {2004})},\ \Eprint
  {http://arxiv.org/abs/astro-ph/0407101} {arXiv:astro-ph/0407101} \BibitemShut
  {NoStop}%
\bibitem [{\citenamefont {Iliadis}\ \emph {et~al.}(2016)\citenamefont
  {Iliadis}, \citenamefont {Anderson}, \citenamefont {Coc}, \citenamefont
  {Timmes},\ and\ \citenamefont {Starrfield}}]{Iliadis:2016vkw}%
  \BibitemOpen
  \bibfield  {author} {\bibinfo {author} {\bibfnamefont {C.}~\bibnamefont
  {Iliadis}}, \bibinfo {author} {\bibfnamefont {K.}~\bibnamefont {Anderson}},
  \bibinfo {author} {\bibfnamefont {A.}~\bibnamefont {Coc}}, \bibinfo {author}
  {\bibfnamefont {F.}~\bibnamefont {Timmes}}, \ and\ \bibinfo {author}
  {\bibfnamefont {S.}~\bibnamefont {Starrfield}},\ }\href {\doibase
  10.3847/0004-637X/831/1/107} {\bibfield  {journal} {\bibinfo  {journal}
  {Astrophys. J.}\ }\textbf {\bibinfo {volume} {831}},\ \bibinfo {pages} {107}
  (\bibinfo {year} {2016})},\ \Eprint {http://arxiv.org/abs/1608.05853}
  {arXiv:1608.05853 [astro-ph.SR]} \BibitemShut {NoStop}%
\bibitem [{\citenamefont {I\~nesta G\'omez}\ \emph {et~al.}(2017)\citenamefont
  {I\~nesta G\'omez}, \citenamefont {Iliadis},\ and\ \citenamefont
  {Coc}}]{InestaGomez:2017eya}%
  \BibitemOpen
  \bibfield  {author} {\bibinfo {author} {\bibfnamefont {A.}~\bibnamefont
  {I\~nesta G\'omez}}, \bibinfo {author} {\bibfnamefont {C.}~\bibnamefont
  {Iliadis}}, \ and\ \bibinfo {author} {\bibfnamefont {A.}~\bibnamefont
  {Coc}},\ }\href {\doibase 10.3847/1538-4357/aa9025} {\bibfield  {journal}
  {\bibinfo  {journal} {Astrophys. J.}\ }\textbf {\bibinfo {volume} {849}},\
  \bibinfo {pages} {134} (\bibinfo {year} {2017})},\ \Eprint
  {http://arxiv.org/abs/1710.01647} {arXiv:1710.01647 [astro-ph.IM]}
  \BibitemShut {NoStop}%
\bibitem [{\citenamefont {de~Souza}\ \emph
  {et~al.}(2019{\natexlab{a}})\citenamefont {de~Souza}, \citenamefont
  {Iliadis},\ and\ \citenamefont {Coc}}]{deSouza:2018gdx}%
  \BibitemOpen
  \bibfield  {author} {\bibinfo {author} {\bibfnamefont {R.~S.}\ \bibnamefont
  {de~Souza}}, \bibinfo {author} {\bibfnamefont {C.}~\bibnamefont {Iliadis}}, \
  and\ \bibinfo {author} {\bibfnamefont {A.}~\bibnamefont {Coc}},\ }\href
  {\doibase 10.3847/1538-4357/aafda9} {\bibfield  {journal} {\bibinfo
  {journal} {Astrophys. J.}\ }\textbf {\bibinfo {volume} {872}},\ \bibinfo
  {pages} {75} (\bibinfo {year} {2019}{\natexlab{a}})},\ \Eprint
  {http://arxiv.org/abs/1809.06966} {arXiv:1809.06966 [astro-ph.IM]}
  \BibitemShut {NoStop}%
\bibitem [{\citenamefont {de~Souza}\ \emph
  {et~al.}(2019{\natexlab{b}})\citenamefont {de~Souza}, \citenamefont {Boston},
  \citenamefont {Coc},\ and\ \citenamefont {Iliadis}}]{deSouza:2019pmr}%
  \BibitemOpen
  \bibfield  {author} {\bibinfo {author} {\bibfnamefont {R.~S.}\ \bibnamefont
  {de~Souza}}, \bibinfo {author} {\bibfnamefont {S.~R.}\ \bibnamefont
  {Boston}}, \bibinfo {author} {\bibfnamefont {A.}~\bibnamefont {Coc}}, \ and\
  \bibinfo {author} {\bibfnamefont {C.}~\bibnamefont {Iliadis}},\ }\href
  {\doibase 10.1103/PhysRevC.99.014619} {\bibfield  {journal} {\bibinfo
  {journal} {Phys. Rev. C}\ }\textbf {\bibinfo {volume} {99}},\ \bibinfo
  {pages} {014619} (\bibinfo {year} {2019}{\natexlab{b}})},\ \Eprint
  {http://arxiv.org/abs/1901.04857} {arXiv:1901.04857 [nucl-th]} \BibitemShut
  {NoStop}%
\bibitem [{\citenamefont {Moscoso}\ \emph {et~al.}(2021)\citenamefont
  {Moscoso}, \citenamefont {de~Souza}, \citenamefont {Coc},\ and\ \citenamefont
  {Iliadis}}]{Moscoso:2021xog}%
  \BibitemOpen
  \bibfield  {author} {\bibinfo {author} {\bibfnamefont {J.}~\bibnamefont
  {Moscoso}}, \bibinfo {author} {\bibfnamefont {R.~S.}\ \bibnamefont
  {de~Souza}}, \bibinfo {author} {\bibfnamefont {A.}~\bibnamefont {Coc}}, \
  and\ \bibinfo {author} {\bibfnamefont {C.}~\bibnamefont {Iliadis}},\ }\href
  {\doibase 10.3847/1538-4357/ac1db0} {\bibfield  {journal} {\bibinfo
  {journal} {Astrophys. J.}\ }\textbf {\bibinfo {volume} {923}},\ \bibinfo
  {pages} {49} (\bibinfo {year} {2021})},\ \Eprint
  {http://arxiv.org/abs/2109.00049} {arXiv:2109.00049 [astro-ph.CO]}
  \BibitemShut {NoStop}%
\bibitem [{\citenamefont {Xu}\ \emph {et~al.}(2013)\citenamefont {Xu},
  \citenamefont {Takahashi}, \citenamefont {Goriely}, \citenamefont {Arnould},
  \citenamefont {Ohta},\ and\ \citenamefont {Utsunomiya}}]{Xu:2013fha}%
  \BibitemOpen
  \bibfield  {author} {\bibinfo {author} {\bibfnamefont {Y.}~\bibnamefont
  {Xu}}, \bibinfo {author} {\bibfnamefont {K.}~\bibnamefont {Takahashi}},
  \bibinfo {author} {\bibfnamefont {S.}~\bibnamefont {Goriely}}, \bibinfo
  {author} {\bibfnamefont {M.}~\bibnamefont {Arnould}}, \bibinfo {author}
  {\bibfnamefont {M.}~\bibnamefont {Ohta}}, \ and\ \bibinfo {author}
  {\bibfnamefont {H.}~\bibnamefont {Utsunomiya}},\ }\href {\doibase
  10.1016/j.nuclphysa.2013.09.007} {\bibfield  {journal} {\bibinfo  {journal}
  {Nucl. Phys. A}\ }\textbf {\bibinfo {volume} {918}},\ \bibinfo {pages} {61}
  (\bibinfo {year} {2013})},\ \Eprint {http://arxiv.org/abs/1310.7099}
  {arXiv:1310.7099 [nucl-th]} \BibitemShut {NoStop}%
\bibitem [{Note2()}]{Note2}%
  \BibitemOpen
  \bibinfo {note} {In particular, for the key reactions $\protect \textrm
  {D}(d,n)^3\protect \textrm {He}$ and $\protect \textrm {D}(d,p)^3\protect
  \textrm {H}$ we either interpolate rates and corresponding uncertainties from
  the NACRE~II numerical database or from the outcome of the Bayesian analysis
  of Ref.~\cite {Iliadis:2016vkw}.}\BibitemShut {Stop}%
\bibitem [{\citenamefont {Aghanim}\ \emph
  {et~al.}(2019{\natexlab{b}})\citenamefont {Aghanim} \emph
  {et~al.}}]{Planck18download}%
  \BibitemOpen
  \bibfield  {author} {\bibinfo {author} {\bibfnamefont {N.}~\bibnamefont
  {Aghanim}} \emph {et~al.} (\bibinfo {collaboration} {Planck}),\ }\href@noop
  {} {}\bibinfo {howpublished}
  {\url{https://wiki.cosmos.esa.int/planck-legacy-archive/index.php/Cosmological_Parameters}}
  (\bibinfo {year} {2019}{\natexlab{b}})\BibitemShut {NoStop}%
\bibitem [{\citenamefont {Coc}\ \emph {et~al.}(2014)\citenamefont {Coc},
  \citenamefont {Uzan},\ and\ \citenamefont {Vangioni}}]{Coc:2014oia}%
  \BibitemOpen
  \bibfield  {author} {\bibinfo {author} {\bibfnamefont {A.}~\bibnamefont
  {Coc}}, \bibinfo {author} {\bibfnamefont {J.-P.}\ \bibnamefont {Uzan}}, \
  and\ \bibinfo {author} {\bibfnamefont {E.}~\bibnamefont {Vangioni}},\ }\href
  {\doibase 10.1088/1475-7516/2014/10/050} {\bibfield  {journal} {\bibinfo
  {journal} {JCAP}\ }\textbf {\bibinfo {volume} {10}},\ \bibinfo {pages} {050}
  (\bibinfo {year} {2014})},\ \Eprint {http://arxiv.org/abs/1403.6694}
  {arXiv:1403.6694 [astro-ph.CO]} \BibitemShut {NoStop}%
\bibitem [{\citenamefont {{Foreman-Mackey}}\ \emph {et~al.}(2013)\citenamefont
  {{Foreman-Mackey}}, \citenamefont {{Hogg}}, \citenamefont {{Lang}},\ and\
  \citenamefont {{Goodman}}}]{emcee2013}%
  \BibitemOpen
  \bibfield  {author} {\bibinfo {author} {\bibfnamefont {D.}~\bibnamefont
  {{Foreman-Mackey}}}, \bibinfo {author} {\bibfnamefont {D.~W.}\ \bibnamefont
  {{Hogg}}}, \bibinfo {author} {\bibfnamefont {D.}~\bibnamefont {{Lang}}}, \
  and\ \bibinfo {author} {\bibfnamefont {J.}~\bibnamefont {{Goodman}}},\ }\href
  {\doibase 10.1086/670067} {\bibfield  {journal} {\bibinfo  {journal}
  {{PASP}}\ }\textbf {\bibinfo {volume} {125}},\ \bibinfo {pages} {306}
  (\bibinfo {year} {2013})},\ \Eprint {http://arxiv.org/abs/1202.3665}
  {arXiv:1202.3665 [astro-ph.IM]} \BibitemShut {NoStop}%
\bibitem [{\citenamefont {{Akaike}}(1974)}]{AkaikeIC}%
  \BibitemOpen
  \bibfield  {author} {\bibinfo {author} {\bibfnamefont {H.}~\bibnamefont
  {{Akaike}}},\ }\href {\doibase 10.1109/TAC.1974.1100705} {\bibfield
  {journal} {\bibinfo  {journal} {IEEE Transactions on Automatic Control}\
  }\textbf {\bibinfo {volume} {19}},\ \bibinfo {pages} {716} (\bibinfo {year}
  {1974})}\BibitemShut {NoStop}%
\bibitem [{\citenamefont {{Gelman}}\ \emph {et~al.}(2013)\citenamefont
  {{Gelman}}, \citenamefont {{Hwang}},\ and\ \citenamefont
  {{Vehtari}}}]{2013arXiv1307.5928G}%
  \BibitemOpen
  \bibfield  {author} {\bibinfo {author} {\bibfnamefont {A.}~\bibnamefont
  {{Gelman}}}, \bibinfo {author} {\bibfnamefont {J.}~\bibnamefont {{Hwang}}}, \
  and\ \bibinfo {author} {\bibfnamefont {A.}~\bibnamefont {{Vehtari}}},\
  }\href@noop {} {\bibfield  {journal} {\bibinfo  {journal} {arXiv e-prints}\
  ,\ \bibinfo {eid} {arXiv:1307.5928}} (\bibinfo {year} {2013})},\ \Eprint
  {http://arxiv.org/abs/1307.5928} {arXiv:1307.5928 [stat.ME]} \BibitemShut
  {NoStop}%
\bibitem [{\citenamefont {Kass}\ and\ \citenamefont
  {Raftery}(1995)}]{BayesFactors}%
  \BibitemOpen
  \bibfield  {author} {\bibinfo {author} {\bibfnamefont {R.~E.}\ \bibnamefont
  {Kass}}\ and\ \bibinfo {author} {\bibfnamefont {A.~E.}\ \bibnamefont
  {Raftery}},\ }\href {\doibase 10.1080/01621459.1995.10476572} {\bibfield
  {journal} {\bibinfo  {journal} {Journal of the American Statistical
  Association}\ }\textbf {\bibinfo {volume} {90}},\ \bibinfo {pages} {773}
  (\bibinfo {year} {1995})},\ \bibinfo {note}
  {\url{http://dx.doi.org/10.1080/01621459.1995.10476572}}\BibitemShut
  {NoStop}%
\bibitem [{\citenamefont {Khlebnikov}\ and\ \citenamefont
  {Shaposhnikov}(1988)}]{Khlebnikov:1988sr}%
  \BibitemOpen
  \bibfield  {author} {\bibinfo {author} {\bibfnamefont {S.~Y.}\ \bibnamefont
  {Khlebnikov}}\ and\ \bibinfo {author} {\bibfnamefont {M.~E.}\ \bibnamefont
  {Shaposhnikov}},\ }\href {\doibase 10.1016/0550-3213(88)90133-2} {\bibfield
  {journal} {\bibinfo  {journal} {Nucl. Phys. B}\ }\textbf {\bibinfo {volume}
  {308}},\ \bibinfo {pages} {885} (\bibinfo {year} {1988})}\BibitemShut
  {NoStop}%
\bibitem [{\citenamefont {Harvey}\ and\ \citenamefont
  {Turner}(1990)}]{Harvey:1990qw}%
  \BibitemOpen
  \bibfield  {author} {\bibinfo {author} {\bibfnamefont {J.~A.}\ \bibnamefont
  {Harvey}}\ and\ \bibinfo {author} {\bibfnamefont {M.~S.}\ \bibnamefont
  {Turner}},\ }\href {\doibase 10.1103/PhysRevD.42.3344} {\bibfield  {journal}
  {\bibinfo  {journal} {Phys. Rev. D}\ }\textbf {\bibinfo {volume} {42}},\
  \bibinfo {pages} {3344} (\bibinfo {year} {1990})}\BibitemShut {NoStop}%
\bibitem [{\citenamefont {March-Russell}\ \emph {et~al.}(1999)\citenamefont
  {March-Russell}, \citenamefont {Murayama},\ and\ \citenamefont
  {Riotto}}]{March-Russell:1999hpw}%
  \BibitemOpen
  \bibfield  {author} {\bibinfo {author} {\bibfnamefont {J.}~\bibnamefont
  {March-Russell}}, \bibinfo {author} {\bibfnamefont {H.}~\bibnamefont
  {Murayama}}, \ and\ \bibinfo {author} {\bibfnamefont {A.}~\bibnamefont
  {Riotto}},\ }\href {\doibase 10.1088/1126-6708/1999/11/015} {\bibfield
  {journal} {\bibinfo  {journal} {JHEP}\ }\textbf {\bibinfo {volume} {11}},\
  \bibinfo {pages} {015} (\bibinfo {year} {1999})},\ \Eprint
  {http://arxiv.org/abs/hep-ph/9908396} {arXiv:hep-ph/9908396} \BibitemShut
  {NoStop}%
\bibitem [{\citenamefont {Dreiner}\ and\ \citenamefont
  {Ross}(1993)}]{Dreiner:1992vm}%
  \BibitemOpen
  \bibfield  {author} {\bibinfo {author} {\bibfnamefont {H.~K.}\ \bibnamefont
  {Dreiner}}\ and\ \bibinfo {author} {\bibfnamefont {G.~G.}\ \bibnamefont
  {Ross}},\ }\href {\doibase 10.1016/0550-3213(93)90579-E} {\bibfield
  {journal} {\bibinfo  {journal} {Nucl. Phys. B}\ }\textbf {\bibinfo {volume}
  {410}},\ \bibinfo {pages} {188} (\bibinfo {year} {1993})},\ \Eprint
  {http://arxiv.org/abs/hep-ph/9207221} {arXiv:hep-ph/9207221} \BibitemShut
  {NoStop}%
\bibitem [{\citenamefont {Shu}\ \emph {et~al.}(2007)\citenamefont {Shu},
  \citenamefont {Tait},\ and\ \citenamefont {Wagner}}]{Shu:2006mm}%
  \BibitemOpen
  \bibfield  {author} {\bibinfo {author} {\bibfnamefont {J.}~\bibnamefont
  {Shu}}, \bibinfo {author} {\bibfnamefont {T.~M.~P.}\ \bibnamefont {Tait}}, \
  and\ \bibinfo {author} {\bibfnamefont {C.~E.~M.}\ \bibnamefont {Wagner}},\
  }\href {\doibase 10.1103/PhysRevD.75.063510} {\bibfield  {journal} {\bibinfo
  {journal} {Phys. Rev. D}\ }\textbf {\bibinfo {volume} {75}},\ \bibinfo
  {pages} {063510} (\bibinfo {year} {2007})},\ \Eprint
  {http://arxiv.org/abs/hep-ph/0610375} {arXiv:hep-ph/0610375} \BibitemShut
  {NoStop}%
\bibitem [{\citenamefont {Gu}(2010)}]{Gu:2010dg}%
  \BibitemOpen
  \bibfield  {author} {\bibinfo {author} {\bibfnamefont {P.-H.}\ \bibnamefont
  {Gu}},\ }\href {\doibase 10.1103/PhysRevD.82.093009} {\bibfield  {journal}
  {\bibinfo  {journal} {Phys. Rev. D}\ }\textbf {\bibinfo {volume} {82}},\
  \bibinfo {pages} {093009} (\bibinfo {year} {2010})},\ \Eprint
  {http://arxiv.org/abs/1005.1632} {arXiv:1005.1632 [hep-ph]} \BibitemShut
  {NoStop}%
\bibitem [{\citenamefont {Affleck}\ and\ \citenamefont
  {Dine}(1985)}]{Affleck:1984fy}%
  \BibitemOpen
  \bibfield  {author} {\bibinfo {author} {\bibfnamefont {I.}~\bibnamefont
  {Affleck}}\ and\ \bibinfo {author} {\bibfnamefont {M.}~\bibnamefont {Dine}},\
  }\href {\doibase 10.1016/0550-3213(85)90021-5} {\bibfield  {journal}
  {\bibinfo  {journal} {Nucl. Phys. B}\ }\textbf {\bibinfo {volume} {249}},\
  \bibinfo {pages} {361} (\bibinfo {year} {1985})}\BibitemShut {NoStop}%
\bibitem [{\citenamefont {Casas}\ \emph {et~al.}(1999)\citenamefont {Casas},
  \citenamefont {Cheng},\ and\ \citenamefont {Gelmini}}]{Casas:1997gx}%
  \BibitemOpen
  \bibfield  {author} {\bibinfo {author} {\bibfnamefont {A.}~\bibnamefont
  {Casas}}, \bibinfo {author} {\bibfnamefont {W.~Y.}\ \bibnamefont {Cheng}}, \
  and\ \bibinfo {author} {\bibfnamefont {G.}~\bibnamefont {Gelmini}},\ }\href
  {\doibase 10.1016/S0550-3213(98)00606-3} {\bibfield  {journal} {\bibinfo
  {journal} {Nucl. Phys. B}\ }\textbf {\bibinfo {volume} {538}},\ \bibinfo
  {pages} {297} (\bibinfo {year} {1999})},\ \Eprint
  {http://arxiv.org/abs/hep-ph/9709289} {arXiv:hep-ph/9709289} \BibitemShut
  {NoStop}%
\bibitem [{\citenamefont {McDonald}(2000)}]{McDonald:1999in}%
  \BibitemOpen
  \bibfield  {author} {\bibinfo {author} {\bibfnamefont {J.}~\bibnamefont
  {McDonald}},\ }\href {\doibase 10.1103/PhysRevLett.84.4798} {\bibfield
  {journal} {\bibinfo  {journal} {Phys. Rev. Lett.}\ }\textbf {\bibinfo
  {volume} {84}},\ \bibinfo {pages} {4798} (\bibinfo {year} {2000})},\ \Eprint
  {http://arxiv.org/abs/hep-ph/9908300} {arXiv:hep-ph/9908300} \BibitemShut
  {NoStop}%
\bibitem [{\citenamefont {Kawasaki}\ and\ \citenamefont
  {Murai}(2022)}]{Kawasaki:2022hvx}%
  \BibitemOpen
  \bibfield  {author} {\bibinfo {author} {\bibfnamefont {M.}~\bibnamefont
  {Kawasaki}}\ and\ \bibinfo {author} {\bibfnamefont {K.}~\bibnamefont
  {Murai}},\ }\href@noop {} {\  (\bibinfo {year} {2022})},\ \Eprint
  {http://arxiv.org/abs/2203.09713} {arXiv:2203.09713 [hep-ph]} \BibitemShut
  {NoStop}%
\end{thebibliography}%

\end{document}